%% file: paper.tex
\documentclass[10pt,conference]{IEEEtran}

\IEEEoverridecommandlockouts
\usepackage{cite}
\usepackage{amsmath,amssymb,amsfonts}
\usepackage{algorithmic}
\usepackage{graphicx}
\usepackage{textcomp}
\usepackage{xcolor}
\usepackage{soul}
\usepackage{tabularx}
\usepackage{tcolorbox}
\usepackage{amsmath}
\usepackage{graphicx}
\usepackage{hyperref}
\usepackage{balance}
\usepackage{makecell}
\usepackage{multirow}
\usepackage{booktabs}
\usepackage{tcolorbox}
\usepackage{threeparttable}
\usepackage{color, colortbl}
\usepackage{subcaption}
\usepackage{xspace}

\newcommand{\sectopic}[1]{\vspace{0.2em}\par\noindent{\textit{\bfseries #1}}}

\def\BibTeX{{\rm B\kern-.05em{\sc i\kern-.025em b}\kern-.08em
    T\kern-.1667em\lower.7ex\hbox{E}\kern-.125emX}}

\begin{document}

\title{AI-based Question Answering Assistance for Analyzing Natural-language Requirements}

\author{
    \IEEEauthorblockN{Saad Ezzini\IEEEauthorrefmark{1}, Sallam Abualhaija\IEEEauthorrefmark{1}, Chetan Arora\IEEEauthorrefmark{3}\IEEEauthorrefmark{4}, Mehrdad Sabetzadeh\IEEEauthorrefmark{2} 
   }
    \IEEEauthorblockA{\IEEEauthorrefmark{1}SnT Centre for Security, Reliability and Trust, University of Luxembourg, Luxembourg}
    \IEEEauthorblockA{\IEEEauthorrefmark{3}Deakin University, Geelong, Australia}
    \IEEEauthorblockA{\IEEEauthorrefmark{4}Monash University, Victoria, Australia}
    \IEEEauthorblockA{\IEEEauthorrefmark{2}School of Electrical Engineering and Computer Science, University of Ottawa, Canada}
    Email: \{saad.ezzini, sallam.abualhaija\}@uni.lu, chetan.arora@monash.edu, m.sabetzadeh@uottawa.ca
}

\maketitle
\begin{abstract}
\input{sections/abstract}
\end{abstract}


\input{sections/introduction}

\input{sections/background}
\input{sections/approach}
\input{sections/evaluation}

\input{sections/threats}
\input{sections/related}
\input{sections/conclusion}
\input{sections/ack.tex}


\bibliographystyle{IEEEtran}
\balance
\bibliography{paper}

\end{document}

%% file: sections/abstract.tex
By virtue of being prevalently written in natural language (NL), requirements are prone to various defects, e.g., inconsistency and incompleteness. As such, requirements are frequently subject to quality assurance processes. These processes, when carried out entirely manually, are tedious and may further overlook important quality issues due to time and budget pressures. In this paper, we propose \textit{QAssist} -- a question-answering (QA) approach that provides automated assistance to stakeholders, including requirements engineers, during the analysis of NL requirements. Posing a question and getting an instant answer is beneficial in various quality-assurance scenarios, e.g., incompleteness detection. 
Answering requirements-related questions automatically is challenging since the scope of the search for answers can go beyond the given requirements specification. To that end, \textit{QAssist} provides support for mining external domain-knowledge resources.  
Our work is one of the first initiatives to bring together QA and external domain knowledge for addressing requirements engineering challenges. We evaluate \textit{QAssist} on a dataset covering three application domains and containing a total of 387 question-answer pairs. We experiment with state-of-the-art QA methods, based primarily on recent large-scale language models. In our empirical study, \textit{QAssist} localizes the answer to a question to three passages within the requirements specification and within the external domain-knowledge resource with an average recall of 90.1\% and 96.5\%, respectively. \textit{QAssist} extracts the actual answer to the posed question with an average accuracy of 84.2\%. 

%% file: sections/introduction.tex
\vspace*{-.05em}
\section{Introduction}\label{sec:introduction}
 
A software requirements specification (SRS) is a pivotal artifact in Requirements Engineering (RE). An SRS lays out
the desired characteristics, functions, and qualities of a proposed system~\cite{vanLamsweerde:09}. 
SRSs are frequently analyzed by requirements engineers as well as by other stakeholders to ensure the quality of the requirements~\cite{Pohl:10}. 
To enable the creation of a shared understanding among stakeholders from different backgrounds, e.g., product managers, domain experts, and developers, requirements are most often written in natural language (NL)~\cite{Zhao:20}. 
Despite its numerous advantages, NL is highly prone to issues such as ambiguity~\cite{Ferrari:19, Ezzini:21}, incompleteness~\cite{Dalpiaz:18,Arora:19} and inconsistency~\cite{Hadar:19}. 
Manually screening for such issues in a lengthy SRS with tens or hundreds of pages is time-consuming, since such screening requires domain knowledge for accurately interpreting the requirements. Evoking domain knowledge is not always quick or easy for humans.  

\begin{figure*}
  \includegraphics[width=0.95\textwidth]{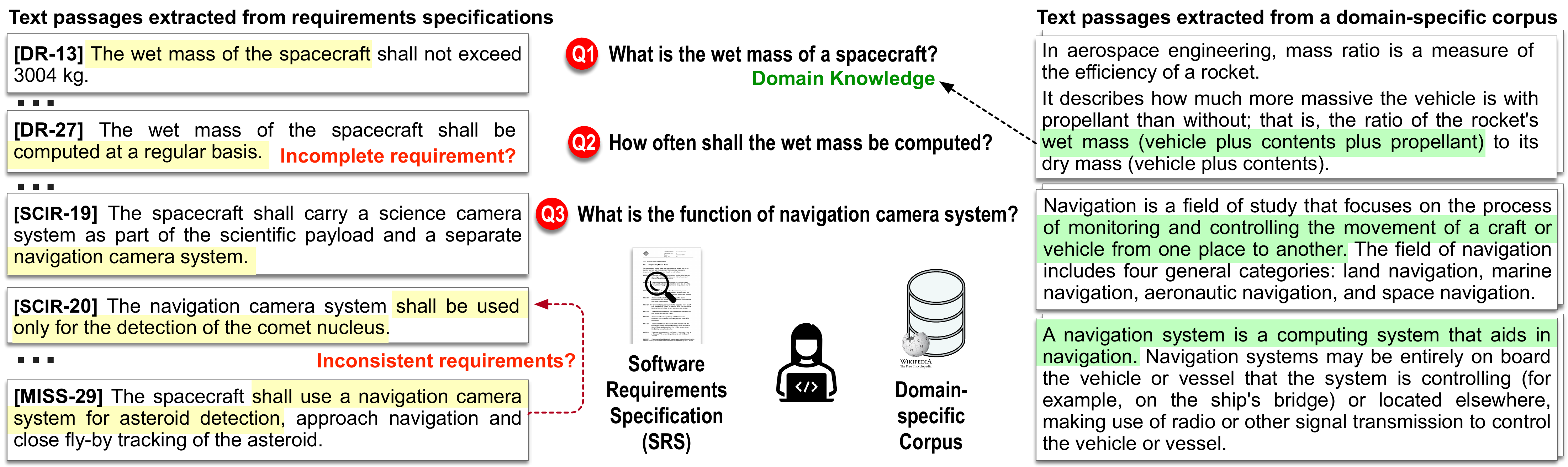}
  \caption{\protect\hbox{Example -- from left to right: passages from SRS; posed questions; passages from external domain-specific corpus.}}
  \label{fig:example}
\end{figure*}

Question answering (QA), i.e., mechanisms to instantly obtain answers to questions posed in NL~\cite{Jurafsky:20}, would be useful as a way to make requirements quality assurance more efficient. 
To illustrate, consider the example requirements in Fig.~\ref{fig:example}. These requirements originate from an SRS in the aerospace domain. To facilitate illustration, the requirements in the figure are prefixed with identifiers. For simplicity, we further assume that each requirement  in our example is one text passage. In practice, a passage, as we elaborate in Section~\ref{sec:approach}, can be made up of multiple consecutive sentences, potentially covering multiple requirements. 

While analyzing requirement \textbf{DR-13} in Fig.~\ref{fig:example}, the developer implementing the computations related to the ``wet mass'' of a spacecraft might come up with question Q1, also shown in Fig.~\ref{fig:example}. Q1 could be prompted by the developer having doubts about the concept of ``wet mass'' or them wanting to verify their interpretation. A challenge here is that the answer to a posed question may be absent from the SRS. This happens to be the case for Q1. Since the presence of a requirements glossary cannot be taken for granted either~\cite{Arora:17}, to answer Q1, one may need to consult external domain resources. These resources could be other SRSs from the same domain, or when such SRSs are non-existent or sparse, a domain-specific corpus extracted from a general source such as Wikipedia. On the right side of Fig.~\ref{fig:example}, we show excerpts of a domain-specific corpus automatically extracted from Wikipedia using an existing open-source corpus extractor~\cite{Ezzini2022wikidominer}. As seen from the figure, just like the SRS being examined, the corpus is made up of passages. These passages may nonetheless be dispersed across multiple documents in the corpus. An answer to Q1 can be found in the extracted corpus. This answer can guide analysts toward making the SRS more complete by providing additional \hbox{information in the SRS about the concept of ``wet mass''.}

In Fig.~\ref{fig:example}, we provide two further questions, Q2 and Q3, that can tip analysts to potential quality problems in our example SRS. Automated QA will find \textbf{DR-27} and more specifically the highlighted segment in that requirement to be an answer to Q2. Upon examining this answer and not finding the exact frequency of wet-mass computations, the analysts will likely conclude that the SRS is incomplete. For a final example, consider Q3. In response to this question, QA identifies several likely answers both in the SRS as well as in the extracted corpus. Among the answers are segments from requirements \textbf{SCIR-20} and \textbf{MISS-29}. Reviewing these two requirements side by side (rather than encountering them potentially many pages apart in the SRS) provides the analysts with a much better chance of noticing the inconsistency between what the two requirements expect of the ``navigation camera system''. The answers from the domain-specific corpus and the passages where these answers are located provide additional useful information for the review process.

In this paper, we propose \textit{QAssist} -- standing for Question Answering Assistance for Improved Requirements Analysis. 
\textit{QAssist} builds on \emph{open-domain QA}, which is the task of finding in a collection of documents the answer to a given question~\cite{Zhu:21}. 
\textit{QAssist} takes as input a question posed in NL and returns as output a list of text passages that likely contain the answer to the question. \textit{QAssist} further demarcates a possible answer (text segment) within each retrieved passage. 
Given questions such as Q1, Q2 and Q3 in Fig.~\ref{fig:example}, we are interested in two sets of text passages: one obtained from the SRS under analysis (left side of Fig.~\ref{fig:example}) and the other obtained by mining a domain-specific knowledge resource (right side of Fig.~\ref{fig:example}). These passages and the answers found within them provide a focused view, helping analysts better understand the requirements and more effectively pinpoint quality problems.

\sectopic{Contributions. }
Our contributions are as follows: 

(1) We devise \textit{QAssist}, an AI-based QA approach aimed at providing assistance with requirements analysis. Given a question posed in NL about the requirements in an SRS, \textit{QAssist} employs Natural Language Processing (NLP) to retrieve two lists of relevant text passages: one from the SRS and one from a domain-specific corpus. In each passage, the likely answer to the posed question is highlighted. When a domain-specific corpus does not exist, \textit{QAssist} automatically builds one, using the phrases appearing in the given SRS as seed terms.
Our implementation of \textit{QAssist} is publicly available~\cite{qassist-rep}.

(2) We develop in a semi-automatic manner a QA dataset tailored to NL requirements. We name this dataset \textit{REQuestA} -- standing for Requirements Engineering Question-Answering dataset. \textit{REQuestA} has been built by two third-party human analysts over six SRSs spanning three application domains. Overall, \textit{REQuestA} contains 387 question-answer pairs. Of these, 214 are manually defined; the remaining 173 are generated automatically and then subjected to manual validation. We make the \textit{REQuestA} dataset publicly available~\cite{qassist-rep}.

(3) We empirically evaluate \textit{QAssist} on the \textit{REQuestA} dataset.  
Our results indicate that \textit{QAssist} retrieves with an accuracy of 100\% from a domain-specific corpus the document that is most relevant to a given question. Furthermore, \textit{QAssist} localizes the answer to a question to three passages within the requirements specification and within the corpus with an average recall of 90.1\% and 96.5\%, respectively. \textit{QAssist} demarcates the actual answer to a question with an average accuracy of 84.2\%.

\sectopic{Significance.} We believe our work is significant for the RE and NLP communities, as we discuss next.
In RE, automated QA has been investigated only to a limited extent and mostly in the context of traceability~\cite{Maletic:09,Mader:13,Pruski:15,Lin:17, Malviya:17}. Traceability QA primarily targets the relationship between different artifacts, e.g., requirements, design diagrams and source code. 
More recently, QA has been studied for improving the understanding of compliance requirements~\cite{Abualhaija:22}. 
%
For RE, the significance of our work is two-fold. First, our QA solution is, to our knowledge, the first to empirically investigate the application of modern QA technologies over industrial requirements. Through a seamless process of posing questions and getting instant answers, our approach enables analysts to explore potential quality issues, e.g., incompleteness and inconsistency. Second, alongside our  solution, we build and publicly release a considerably sized QA dataset covering six SRSs from three application domains. This dataset focuses on clarification questions posed over SRSs and is the first dataset of its kind.

QA is widely studied in the NLP community~\cite{Soares:20}, where, as we elaborate in Section~\ref{sec:related}, many automated solutions and datasets have been proposed and evaluated. The most well-known QA datasets in the NLP literature are derived from Wikipedia, e.g., SQUAD~\cite{SQUAD:16}, TriviaQA~\cite{Joshi:17} and NQ~\cite{Kwiatkowski:19}, to name few. There are also examples of domain-specific datasets, e.g., in the medical~\cite{Pampari:18,He:19,Tian:19} and railway~\cite{Hu:20} domains.
From an NLP standpoint, our work is significant in that it is capable of looking beyond a single source for identifying answers to a posed question.
The NLP literature concentrates mainly on the situation where the answer to a question resides in an a-priori-known source document (or text passage). Our work departs from this position by bringing in a secondary source of knowledge, namely a domain-specific corpus, to complement the primary source (in our case, an SRS), while maintaining the distinction between the two sources. Using a secondary source is necessitated by our application context: SRSs are typically highly technical with a potentially large amount of tacit (unstated) domain knowledge underpinning them. By provisioning for, and when necessary, automatically constructing a domain-specific corpus, our approach increases the chance that analysts will obtain satisfactory answers to \hbox{their requirements-related questions.}

\sectopic{Structure. }
Section~\ref{sec:background} presents background.  Section~\ref{sec:approach} describes our QA approach.
Section~\ref{sec:evaluation} reports on our empirical evaluation. \textcolor{black}{Section~\ref{sec:google} compares with broad-based search engines. }
Section~\ref{sec:threats} explores threats to validity. 
Section~\ref{sec:related} discusses related work. Section~\ref{sec:conclusion} concludes the paper.

%% file: sections/background.tex
\section{Background}\label{sec:background}

This section describes the background for our QA approach. 

\sectopic{Open-domain QA.} Our proposed approach targets the open-domain QA task (defined in Section~\ref{sec:introduction}). Modern open-domain QA  solutions work in two stages, combining information retrieval (IR) with  machine reading comprehension (MRC)~\cite{Chen:17}.
IR is applied first to narrow the search space by finding the relevant text passages that likely contain the answer to a question~\cite{McGill:83}.
Subsequently, MRC models extract the likely answer to the question from the text passages retrieved~\cite{SQUAD:16}. 
An IR-based method is referred to as a  \textsc{retriever} since it retrieves relevant text, while an MRC-based model is referred to as a \textsc{reader} since it reads the text to find the answer~\cite{Zhu:21}.
State-of-the-art QA techniques rely heavily on language models (LMs) such as BERT~\cite{Devlin:18} as an enabling technology~\cite{Liu:19}. %
Below, we introduce IR and MRC alongside the LMs that we consider and experiment with in \hbox{the development of our approach.}
\sectopic{Information Retrieval (IR). } Given a query and a collection of documents, IR methods are designed to rank the documents according to their relevance to the query~\cite{Manning:08}. 
Traditional methods in IR include term frequency - inverse document frequency (TF-IDF) and Okapi Best Matching (BM25). 
TF-IDF assigns a composite weight to each term occurring in the document depending on its occurrence frequency in the document relative to its frequency in the entire document collection~\cite{Jones:72}. These weights are used to transform a text sequence into a mathematical vector.
Following this, both the query and the documents are represented as vectors, with the query being treated as a (short) document. Relevance is computed using similarity metrics, e.g., cosine similarity~\cite{Manning:08}. Similarity metrics quantify the similarity between the query and a document while normalizing the difference in vector length; vectors for documents are significantly longer than those for queries. 
Unlike TF-IDF which is a binary model relying on the presence of question terms in the document collection, BM25 is a probabilistic model that improves the TF-IDF weights using relevance feedback~\cite{Robertson:09}. 

In the context of QA, IR-based methods assess relevance of documents as well as text passages within individual documents. In the latter case, each passage is regarded as a single document during vectorization.
Despite being relatively old, BM25 and to a lesser extent TF-IDF are still widely applied in  text retrieval tasks due to their simple implementation and robust behavior~\cite{Thakur:21}. 
In addition to traditional methods, dense and reranking methods have recently been introduced  in the QA literature~\cite{Nogueira:19, Thakur:21,Wang:21,Zhuang:21}. Leveraging language models, dense methods compute relevance based on the text representations in the dense vector space, whereas reranking methods combine the rankings of two different IR-based methods. 

\sectopic{Machine Reading Comprehension (MRC).} MRC models are specifically used to extract the likely answer to a given question from a text passage~\cite{SQUAD:16}. 
MRC is often solved using pre-trained language models (e.g., BERT), introduced next. These models typically limit  the length of the text  passage to be less than or equal to 512 tokens~\cite{Devlin:18,chen:20}.

\sectopic{Language Models (LMs).} Large-scale neural LMs have rapidly  dominated the state-of-the-art in NLP~\cite{Lewis:20}. LMs are pre-trained on large  bodies of text in order to learn contextual information, regularities of language, and syntactic and semantic relations between words. This learned knowledge can then be used by fine-tuning LMs to solve downstream NLP tasks~\cite{Pan:09}, e.g., QA~\cite{Petroni:19}. Below, we briefly discuss the LMs that we consider and experiment with in this paper.

\textit{Bidirectional Encoder Representations from Transformers (BERT)~\cite{Devlin:18}} is pre-trained on the BooksCorpus and English Wikipedia with two training objectives, namely masked language modeling (MLM) and next sentence prediction (NSP). 
In MLM, a fraction of the tokens in the pre-training text are randomly masked. The model is trained to predict the original vocabulary of these masked tokens based on the surrounding context. For example, BERT should predict the masked token ``briefed'' in the phrase ``\texttt{[MASK]} reporters on''. In NSP, the model is trained to predict whether two text segments are consecutive in the original text. 
BERT learns contextualized representations of words by utilizing the \emph{Transformer} architecture~\cite{Vaswani:17} and attention mechanisms that allow the model to attend to different information from different representations. For example, the model re-weights the embeddings of ``bank'' and ``river'' in the sentence ``I walked along the banks of the river'' to highlight the meaning of ``bank'' in this context.

Efficiently Learning an Encoder that Classifies Token Replacements Accurately (\textit{ELECTRA})~\cite{Clark:20} improves the contextual representations learned by BERT by replacing the MLM training objective with a token replacement objective, i.e., randomly replacing some tokens instead of masking them. 

A Lite BERT (\textit{ALBERT})~\cite{Lan:19}, A Distilled Version of BERT (\textit{DistilBERT})~\cite{Sanh:19}, \textit{MiniLM}~\cite{Wang:20a} and the Robustly optimized BERT pre-training approach (\textit{RoBERTa})~\cite{Liu:19a} are other variants that optimize the size and computational cost of BERT using methods such as knowledge distillation~\cite{Gou:21} -- a technique that transfers knowledge from a large unwieldy model to generate a smaller model with less parameters yet similar performance.

\begin{figure*}[!t]
\centering
  \includegraphics[width=0.95\textwidth]{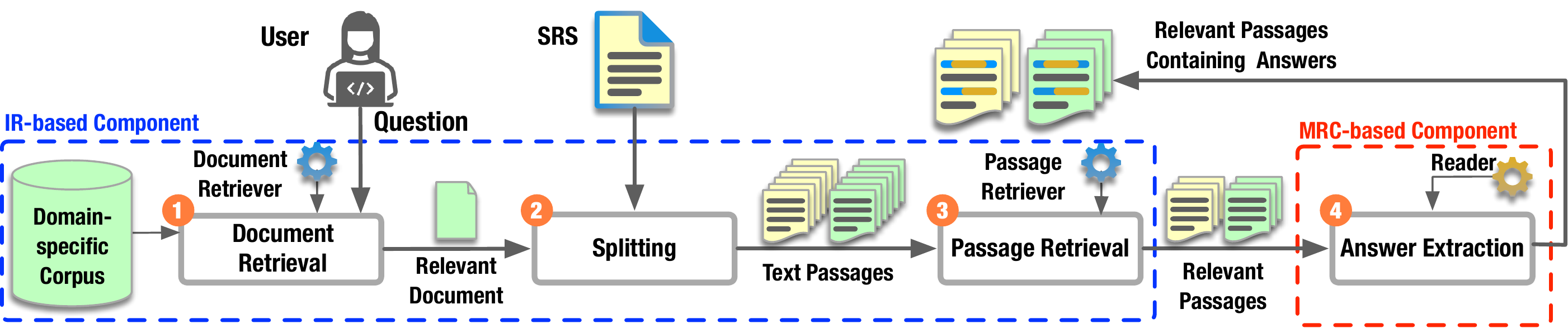}
  \caption{Overview of our approach (\textit{QAssist}).}
  \label{fig:approach}
\end{figure*}

The text-to-text transfer transformer (\textit{T5}) model~\cite{Raffel:19} is another interesting and popular LM. T5 is pre-trained on the Colossal Clean Crawled Corpus (C4) which was also released alongside the model. C4 consists of hundreds of gigabytes of clean text that is crawled from the Web. Compared to BERT-style models, T5 uses a text-to-text framework that enables addressing a wider spectrum of NLP downstream tasks as long as they can be formulated as a text-to-text problem.

%% file: sections/approach.tex
\section{Approach}\label{sec:approach}

In this section, we describe our approach and also establish the notation that we use throughout the rest of the paper. Fig.~\ref{fig:approach} shows an overview of our approach.
\textit{QAssist} takes as input a question ($q$) posed in NL by the user and an SRS. 
In step 1, \textit{QAssist} retrieves the most relevant document ($d$) to $q$ from an external domain-specific corpus ($\mathcal{D}$).
%
In step~2, \textit{QAssist} generates a list of text passages ($\mathcal{T}$) by splitting the input SRS and $d$. 
\textit{QAssist} then finds in step~3 the top-$k$  text passages ($\mathcal{R} \subset \mathcal{T}$) that are most relevant to $q$. 
In step~4, \textit{QAssist} extracts a likely answer from each text passage retrieved in step~3. 
\textit{QAssist} finally returns as output the relevant text passages from step~3 alongside the answers extracted in step~4. 
%
As explained in Section~\ref{sec:background}, the pipeline for an open-domain QA system like \textit{QAssist} is made up of two phases: (i)~IR-based (spanning steps~1 -- 3) and (ii)~MRC-based (step~4).
In phase~(i), we apply \emph{two} \textsc{retrievers}, one for retrieving $d \in \mathcal{D}$ in step~1 (\textit{document retriever} -- for short \textsc{retriever}$_D$) and another for finding $\mathcal{R}$ in step~3 (\textit{passage retriever} -- for short \textsc{retriever}$_T$). 
Next, we elaborate each step of \textit{QAssist}. 

\subsection{Step~1: Document Retrieval}\label{subsec:dRetrieval}
As a prerequisite for applying \textsc{retriever}$_D$ in this step, a corpus $\mathcal{D}$ should be available. 
When $\mathcal{D}$ is absent, it can be automatically generated using existing corpus-extraction methods~\cite{Milne:06,Cui:08,ferrari:17,Ezzini:21,Saxena:21,Ezzini2022wikidominer}. \textit{QAssist}'s ability to incorporate an external corpus of knowledge into the QA process is important as a way to enrich the output with domain knowledge. 
In this step, \textsc{retriever}$_D$ mines $\mathcal{D}$ to find a document $d$ that is most relevant to $q$. 
In particular, \textsc{retriever}$_D$ computes first the relevance between $q$ and each document in $\mathcal{D}$, and then ranks these documents according to relevance scores. 
From the resulting ranked list, \textit{QAssist} selects as the result of step~1 the \hbox{most relevant document ($d \in \mathcal{D}$).}
%
Note that, while unnecessary for our purposes in this paper, the number of most relevant documents to retrieve from $\mathcal{D}$ can be configured to a value $c > 1$. In that case, the output $d$ from step~1 would be the sequential combination of the top-$c$ retrieved documents.

\subsection{Step~2: Splitting}\label{subsec:partitioning} 
This step takes two documents as input: the SRS under analysis as well as the most relevant corpus document $d$ retrieved in step~1. \textit{QAssist} automatically generates two lists $\mathcal{T}_S$ and $\mathcal{T}_D$ of text passages by splitting the given SRS~and~$d$, respectively. 
To do so, we employ a simple NLP pipeline that consists of \textit{tokenization} and \textit{sentence splitting}, breaking the input text (SRS and $d$) into tokens and sentences. 
Using the annotations from this NLP pipeline, we iterate over each document to identify the text passages. 

Recall from Section~\ref{sec:background} that LM-based \textsc{readers} (which we apply in step~4) typically limit passage length to 512 tokens. 
\textcolor{black}{Accordingly, we define a \textit{text passage} as a paragraph, unless the paragraph is too long (i.e., has more than 512 tokens) and thus cannot be processed by LMs in its entirety. Long paragraphs are split with one sentence of overlap to preserve context. 
} 
\textcolor{black}{Concretely, we} apply the following procedure to split \textcolor{black}{long paragraphs} into coherent passages. 

Assume that a given paragraph has a sequence of $n$  sentences, $s_1, \ldots, s_n$. We put consecutive sentences $s_1, \ldots, s_i$ into one passage, such that the length of the resulting passage is less than or equal to 512 tokens. In the next iteration, we start at $s_i$, i.e., the last sentence of the previous passage. To create the next passage, we take consecutive sentences $s_i, \ldots, s_j$ subject to the 512-token length constraint. This process is repeated until all the sentences in the paragraph have been covered.
The rationale for a one-sentence overlap between adjacent passages from the same paragraph is to help maintain flow continuity in the passages.  
%

The output from step~2 ($\mathcal{T}_S$ and $\mathcal{T}_D$) is passed to step~3. 

\subsection{Step~3: Passage Retrieval}\label{subsec:cRetrieval}
In this step, we apply \textsc{retriever}$_T$ to find the $k$ most relevant text passages to $q$ from each $\mathcal{T}_S$ and $\mathcal{T}_D$. We denote the set of resulting passages by $\mathcal{R}_S \subset \mathcal{T}_S$ and $\mathcal{R}_D \subset \mathcal{T}_D$, respectively. 
In a similar manner to step~1, \textsc{retriever}$_T$ computes and assigns relevance scores to each text passage in $\mathcal{T}_S$ and $\mathcal{T}_D$. The passages in each $\mathcal{T}_S$ and $\mathcal{T}_D$ are sorted in descending order of relevance  and the top-$k$ passages are picked. 
In Section~\ref{sec:evaluation}, we empirically assess the implications of the value of $k$ for practice. 
$\mathcal{R}_S$ and \hbox{$\mathcal{R}_D$ constitute the input to step~4.}

\subsection{Step~4: Answer Extraction}\label{subsec:answer}
In the last step of \textit{QAssist}, we apply a \textsc{reader} to extract a likely answer to $q$ from each text passage in $\mathcal{R}_S$ and $\mathcal{R}_D$. The likely answers are highlighted in and presented together with $\mathcal{R}_S$ and $\mathcal{R}_D$ as the output of \textit{QAssist}.
Which \textsc{reader} technology yields the best results is a question that we  investigate empirically in Section~\ref{sec:evaluation}.

%% file: sections/evaluation.tex
\section{Empirical Evaluation} \label{sec:evaluation}

In this section, we empirically evaluate \textit{QAssist}. 

\subsection{Research Questions (RQs)}

Our evaluation addresses the following RQs:

\sectopic{RQ1: Which \textsc{retriever} has the highest accuracy in finding text that is most relevant to a given question?}
Recall from Section~\ref{sec:approach} that \textit{QAssist} employs \textsc{retriever}$_D$ in step~1 (i.e., \textit{Document Retrieval}) and \textsc{retriever}$_T$ in step~3 (i.e., \textit{Passage Retrieval}). \textsc{Retriever}$_D$ takes as input a collection of documents and returns as output the most relevant document $d\in\mathcal{D}$. \textsc{Retriever}$_T$ takes as input a list of text passages and return as output the top-$k$ passages relevant to  a given question. For each \textsc{retriever}, we investigate in RQ1 four alternatives  from the IR literature as outlined in Section~\ref{subsec:evaluationProc}.  
RQ1 identifies the most accurate alternative for each \textsc{retriever}.

\sectopic{RQ2: Which \textsc{reader} produces the most accurate results for extracting the likely answer to a given question? }
\textit{QAssist} uses in step~4 (i.e., \textit{Answer Extraction}) a \textsc{reader} for extracting a likely answer to a given question from each relevant text passage retrieved by the \textit{passage retrievers} in step~3. 
Multiple alternative \textsc{readers} can be applied here as we explain in Section~\ref{subsec:evaluationProc}. RQ2 investigates these alternatives and identifies the most accurate one.

\sectopic{RQ3: Does \textit{QAssist} run in practical time?} 
RQ3 analyzes \textit{QAssist}'s execution time. 
To be applicable in practice, \textit{QAssist} needs to be able to answer questions in practical time.

\subsection{Implementation}\label{subsec:implementation}

We implement \textit{QAssist} using Python 3.7.13 and 
Jupyter Notebooks~\cite{Kluyver:16}. 
Specifically, we implement the NLP pipeline (including the tokenizer and sentence splitter) using the Transformers 3.0.1 library~\cite{transformers}. We implement the traditional IR methods and TF-IDF vectorization using Scikit-learn 1.0.2~\cite{scikit-learn}, and implement BM25 using the BM25 0.2.2 library~\cite{rank-bm25}. 
The language models that we experiment with include the IR-based models \textit{DistilBERT-base-tas-b} and \textit{MiniLM-L-12-v2} from BeIR~\cite{beir} and the MRC-based models  
\textit{ALBERT-large v1.0}, \textit{BERT-large-uncased},  \textit{DistilBERT-base-cased}, \textit{ELECTRA-base}, \textit{MiniLM-uncased} and \textit{RoBERTa-base} from HuggingFace~\cite{huggingface}. 
For corpus extraction from Wikipedia, we use the Wikipedia 1.4.0 library~\cite{wikipy}.
For question generation, discussed in Section~\ref{subsec:data_collection}, we use NLTK 3.2.5~\cite{NLTK} to preprocess text from SRSs and corpus documents.  
We then apply \textit{T5-base-question-generator} and \textit{BERT-base-cased-qa-evaluator} for automatically generating and assessing question-answer pairs. Both of these models are from HuggingFace.

\input{subsections/data_collection}

\input{subsections/evaluation_procedure}

\input{subsections/RQs}

%% file: subsections/data_collection.tex
\subsection{Data Collection Procedure}\label{subsec:data_collection}

To evaluate  \textit{QAssist}, we collected six SRSs from three application domains, namely aerospace, defence, and security. 
Our data collection resulted in a QA dataset named \textit{REQuestA} (RE Question-Answering dataset).
To reduce the cost and effort required for the construction of this dataset, about half of the question-answer pairs in \textit{REQuestA} were generated automatically using text generation models~\cite{Raffel:19} and then validated by human analysts. The remaining half were defined manually. In this section, we discuss the desiderata for \textit{REQuestA}, the automatic QA generation method, the process for manual definition of question-answer pairs, and finally the details of the resulting dataset. 

\sectopic{Desiderata. }
We identify the following desiderata for \textit{REQuestA} in view of the analytical goals we would like to support, as discussed in Section~\ref{sec:introduction}. 

\noindent (1) \textit{Focus on content-based questions}. \textit{REQuestA} is populated with clarification questions over SRSs. \textit{REQuestA} thereby does not contain questions that are not directly related to the SRS content, for instance, questions related to change impact analysis or project management, an example of which would be ``How many requirements are not implemented in Phase-1 of the project?''. Questions of this nature are legitimate in RE~\cite{Malviya:17}, but are outside the scope of our current work.

\noindent (2) \textit{Inclusion of external sources of knowledge}. Motivated by covering the domain knowledge that is often left tacit in SRSs, we would like \textit{REQuestA} to include relevant text passages not only from SRSs but also from external sources of knowledge. The inclusion of external knowledge sources enables us to more conclusively evaluate the effectiveness of QA by considering requirements-related questions that would go unanswered based on the contents of a given SRS alone.

\sectopic{QA Auto-generation. } 
Despite the availability of QA datasets, none of them are directly applicable in our work, as explained in Section~\ref{sec:introduction}. Building a ground truth for QA requires considerable manual effort for proposing both questions and answers. This prompted us to consider question generation (QG)~\cite{Du:17,Pan:19} as an aid during dataset construction.
 QG enables automated derivation of a large number of questions and answers from a given knowledge source; these questions and answers can subsequently be subjected to manual validation for correctness. Such validation generally takes less time and cognitive effort from humans than deriving questions and answers from scratch.

\begin{figure}[!t]
\centering
  \includegraphics[width=0.5\textwidth]{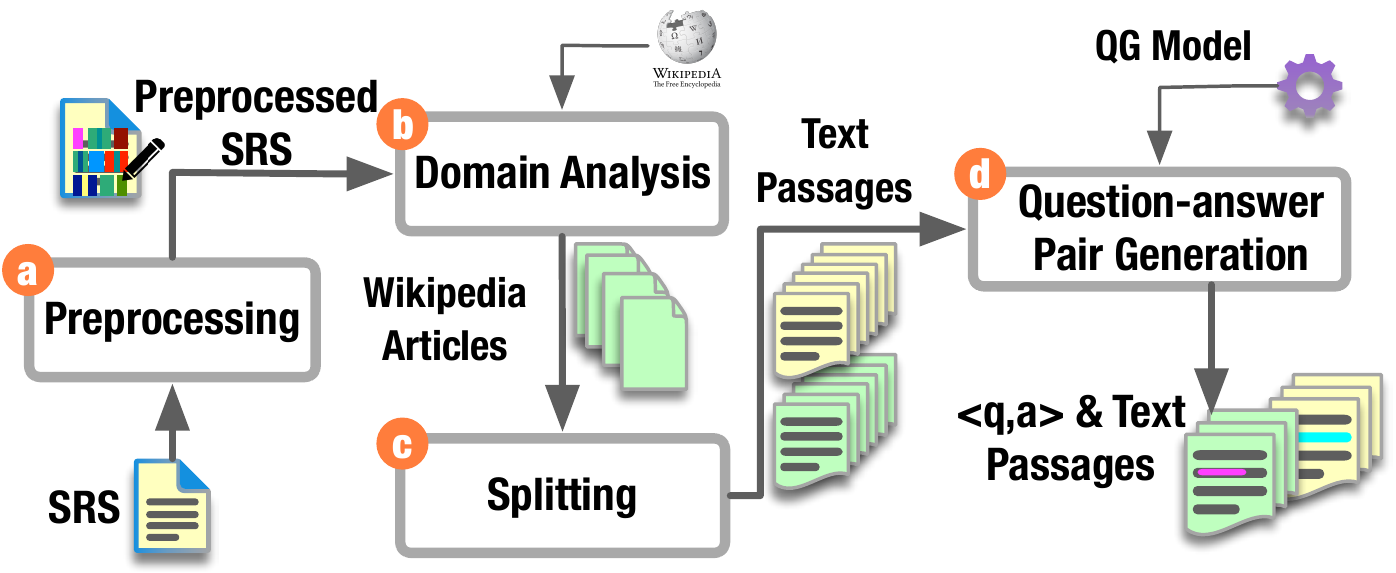}
  \caption{Overview of our question generation method (used exclusively for building our dataset, \textit{REQuestA}).}
  \label{fig:qg-overview}
\end{figure}

An entry in \textit{REQuestA} is  a text passage and a question-answer pair associated with that passage. 
\textit{An answer} in our work is a short text span in a sentence. 
The questions and answers in \textit{REQuestA} are derived from two different sources: the input SRS and a domain-specific corpus created automatically around the content of the input SRS. 
Fig.~\ref{fig:qg-overview} shows an overview of our method for automatically generating questions and answers. Given an SRS as input, our method returns a list of question-answer pairs in four steps, elaborated next.  

\vspace*{.2em}
\noindent\textit{(a) Preprocessing}: In this step, we preprocess the input SRS by applying an NLP pipeline. The goal of this step is to identify a set of concepts which are used in the next step to analyze the domain of the input SRS. To find these concepts, we applied REGICE~\cite{Arora:17} -- an existing tool for extracting  glossary terms from NL requirements. 

\vspace*{.2em}
\noindent \textit{(b) Domain Analysis}: We build in this step a minimal domain-specific corpus. 
To do so, we first group the SRSs from the same domain and then use the concepts extracted from these SRSs in step~(a). 
Specifically, we compute for each concept a TF-IDF score, adapted to work over phrases (e.g., ``navigation camera'') rather than only individual terms (e.g., ``camera''). 
Next, we attempt to increase the specificity of the concepts by removing any generic concepts (e.g., ``camera'') appearing in WordNet~\cite{Miller:95} -- a generic lexical database for English.  
We then sort the concepts in descending order of TF-IDF scores and select the top-50 concepts, referring to these concepts as \textit{keywords}. 
Inspired by recent work on the generation of domain-specific corpora for requirements analysis tasks~\cite{Ezzini:21}, 
we use each keyword to query Wikipedia and find a matching article, i.e., an article whose title overlaps with a given keyword. 
Finally, we randomly select from the matching articles a subset to use in the next step. 

\vspace*{.2em}
\noindent \textit{(c) Splitting}: 
In this step, we use the same method presented in Section~\ref{sec:approach} to automatically split the SRS and Wikipedia articles into a set of text passages.  

\vspace*{.2em}
\noindent \textit{(d) Question-answer Pair Generation}: 
In this step, we use a QG model based on the T5 language model (introduced in Section~\ref{sec:background}). We give as input a text passage to the QG model. The model first extracts a random answer from the passage and then automatically generates a corresponding question. 
For example, for passage \textbf{DR-13} in Fig.~\ref{fig:example}, the QG model could first pick  ``3004 kg'', and then generate the following question: \textit{``What shall the wet mass of the spacecraft not exceed?''}.
The output of the QG model includes the text passage and a  set of automatically generated question-answer pairs. Each such pair will be denoted $\langle q,a\rangle$ hereafter. Note that multiple pairs can be generated from the same text passage.
To reduce the manual effort needed for validating the questions and answers, we apply a QA evaluator that is based on BERT. The evaluator takes as input a pair $\langle q,a\rangle$ and returns as output a value representing its prediction about whether the pair  is valid. 
We sort the auto-generated pairs according to the resulting scores from the evaluator, and then select the top 5\% of the $\langle q,a\rangle$ pairs automatically generated from each SRS and the Wikipedia articles in the respective corpus.

\sectopic{Construction of \textit{REQuestA}.} 
The construction of \textit{REQuestA} involved two third-party (non-author) human analysts.
The first analyst has a Master's degree in multilingualism. The second analyst has  a computer science background with a Master's degree in quality assurance. Both analysts had prior experience with software requirements and had previously contributed to annotation tasks involving SRSs. Before starting their work, the analysts participated in a half-day training session on question answering in RE where they additionally received instructions about the desiderata for \textit{REQuestA}. 

We shared with the analysts the original SRSs, the randomly selected Wikipedia articles (created during the domain analysis step in Fig.~\ref{fig:qg-overview}), and 
the list of automatically generated $\langle q,a\rangle$ pairs for each SRS. 
The analysts were asked to handle each $\langle q,a\rangle$ pair as follows. 
Each question $q$ was labeled as \textit{valid} indicating that $q$ was correct as-is, \textit{rephrased} indicating that $q$ was semantically correct but required structural improvement to become valid, or \textit{invalid} indicating that $q$ did not make  sense. 
Similarly, each answer $a$ was labeled as \textit{correct}, \textit{corrected}, or \textit{invalid} with similar indications to the ones mentioned above for $q$. Additionally, $a$ could be labeled as \textit{not in context} indicating that the question cannot be answered from the given text passage. In this case, we consider the answers as \textit{invalid}.
%
We further asked the analysts to manually define question-answer pairs on each text passage during the validation process. We discuss quality considerations for our dataset later in this section.

To construct the \textit{REQuestA} dataset, we filtered out any pair where either $q$ or $a$ was invalid. For the remaining pairs, we used the rephrased $q$ and corrected $a$ according to the revisions suggested by the human analysts. 
In total, we automatically generated 204 $\langle q,a\rangle$ pairs; 111 from the SRSs and 93 from the Wikipedia articles. 
From these, we filtered 31 pairs due to invalid questions or answers, leaving 173 pairs in the dataset (86 from the SRSs and 87 from the Wikipedia articles). We further included in \textit{REQuestA} question-answer pairs that the analysts had defined manually during the validation process alongside the respective text passages. In total, the analysts manually defined 214 pairs (103 from the SRSs and 111 from the Wikipedia articles). 
Overall, \textit{REQuestA} contains 387 pairs.

Table~\ref{tab:docCollection} provides summary statistics for \textit{REQuestA}.
Specifically, the table lists the number of auto-generated $\langle q,a\rangle$ pairs (\textit{auto}) as well as the number of pairs manually defined by the analysts (\textit{man}).  The table further shows $\overline{|\mathcal{T}_D|}$ indicating the average 
number of text passages in the Wikipedia articles (noting that there are multiple articles in each corpus), and $|\mathcal{T}_S|$ indicating the number of text passages in each SRS.

\input{tables/doc-col}

\sectopic{Quality of \textit{REQuestA}. } As a quality measure, the two analysts reviewed an overlapping subset amounting to 10\% of the auto-generated $\langle q,a\rangle$ pairs. We counted an agreement when the analysts selected the same label for a given question or answer (i.e., valid or invalid), noting that valid includes both rephrased and corrected.
On this subset, the analysts were in full agreement (i.e., no disagreements) on the labels for the questions and answers.

To further ensure the quality of the dataset, we analyzed all the automatically generated questions and answers against the corrections provided by the human analysts.
Out of the 173 valid questions, the analysts collectively rephrased 24 questions (representing $\approx$14\% of the auto-generated questions) and corrected 46 answers (representing $\approx$26\% of the auto-extracted answers).
Out of the 46 corrected answers, 26  were expanded by the analysts to include missing tokens, e.g., the auto-extracted answer ``software code'' was corrected to ``implemented software code''. To increase the quality of our dataset, we included in \textit{REQuestA} the corrected answers and not the auto-extracted ones. 
Following best practices in the natural-language generation literature and machine translation~\cite{Hanna:21}, we apply BLEU for lexical similarity and BERTScore for semantic similarity. 
Given two questions, $q_1$ and $q_2$, BLEU measures the overlapping tokens between $q_1$ and $q_2$. The score is then normalized by the total number of the tokens in $q_1$ and $q_2$. BERTScore measures semantic similarity between $q_1$ and $q_2$ based on contextual word embeddings. 
The resulting scores are BLEU$=$0.54 and BERTScore$=$0.95. These values indicate that the auto-generated questions and the rephrased ones are semantically very similar albeit using different structures. 
These scores indicate that our QG method successfully produces semantically correct questions, while also implying that the analysts frequently chose to make structural improvements for better readability. 

Since no training or fine-tuning is performed in our approach, we use \textit{REQuestA} in its entirety for empirically evaluating the available QA technologies.
To facilitate replication and future research, \textit{REQuestA} is made publicly available~\cite{qassist-rep}.

%% file: tables/doc-col.tex
\begin{table}
\caption{Summary Statistics for the \textit{REQuestA} Dataset.} %
\label{tab:docCollection}
\begin{threeparttable}[t]
\centering
\begin{tabularx}{0.49\textwidth}{@{}l*{7}{>{\centering\arraybackslash}X}@{}}
\toprule
&& \multicolumn{2}{c}{$\langle q,a\rangle$} &&& \multicolumn{2}{c}{$\langle q,a\rangle$}   \\
\cmidrule(lr){3-4}\cmidrule(lr){7-8}
Domain &  $\overline{|\mathcal{T}_D|}$ &\textit{auto}& \textit{man} & SRS & $|\mathcal{T}_S|$ & \textit{auto} & \textit{man}  \\
\midrule
\multirow{2}{*}{Aerospace}  & \multirow{2}{*}{42} &\multirow{2}{*}{45}  & \multirow{2}{*}{53} & \#1  & 24  & 8  & 18   \\ 
&&&&\#2  & 107  & 37 & 40\\ 
\midrule
\multirow{2}{*}{Defence} &  \multirow{2}{*}{94} &\multirow{2}{*}{38}  & \multirow{2}{*}{50} & \#3 & 11  & 5  & 4 \\    
&&&&\#4 & 71  & 19 & 26 \\ 
\midrule
\multirow{2}{*}{Security} &  \multirow{2}{*}{23} &\multirow{2}{*}{4}  & \multirow{2}{*}{8} &\#5& 18  & 15 & 13 \\
&&&&\#6& 4   & 2 & 2  \\ 
\midrule
Total & 159 & 87 & 111 & - & 235 & 86 & 103  \\
\bottomrule
\end{tabularx}
 \end{threeparttable}
 \vspace*{-.8em}
\end{table}

%% file: subsections/evaluation_procedure.tex
\subsection{Evaluation Procedure} \label{subsec:evaluationProc}

To answer our RQs, we conduct the following experiments. See Section~\ref{sec:background} for background.

\sectopic{EXPI.} This experiment answers \textbf{RQ1}. We evaluate in EXPI four alternative \textsc{retrievers}, including the traditional \textsc{retrievers} TF-IDF and BM25, DistilBERT dense \textsc{retriever}, and a reranking \textsc{retriever} that pairs BM25 with MiniLM cross encoder. 
We identify in EXPI the most accurate \textsc{retriever} applied in step~1 of our approach (Fig.~\ref{fig:approach}) for retrieving the most relevant external document from a domain-specific corpus. 
We further identify the most accurate \textsc{retriever} in step~3 for retrieving from the input SRS and the most relevant external document the top-$k$ relevant text passages for a given question.
We compare the performance of the alternative \textsc{retrievers} using two evaluation metrics commonly used in the IR literature~\cite{McGill:83}.  
The first metric is \textit{recall@k (R@$k$)} and assesses whether the document (or text passage) containing the correct answer to a given question ($q$) is in the ranked list of the top-$k$ documents (or passages) produced by the \textsc{retriever}.
The second metric, \textit{normalized discounted cumulative gain@k (nDCG@$k$)}, is similar to \textit{R@$k$}, except that it accounts not only for the mere presence of the relevant document (or passage) but also for its rank. 

We note that we are interested only in the most relevant document (top-$1$) retrieved by the document \textsc{retriever} 
in step~1 of our approach. In this case, ranking is not relevant and the above two metrics produce the same result; we thus report only R@$1$ for the document \textsc{retriever}. 
To run EXPI, using an existing open-source tool~\cite{Ezzini2022wikidominer}, we generate domain-specific corpora covering the aerospace, defence, and security domains and corresponding to the SRSs in our study.   

\vspace*{.2em}\sectopic{EXPII.} 
This experiment answers \textbf{RQ2}. To extract the answer to a given question in step~4 of our approach (Fig.~\ref{fig:approach}), we experiment with the following alternative \textsc{readers}: 
ALBERT, BERT, DistilBERT, ELECTRA, MiniLM, and RoBERTa.
We compare the performance of the \textsc{readers} using \textit{Accuracy (A)}, computed as the number of questions correctly answered by the \textsc{reader} divided by the total number of questions. 
To decide whether an answer is correct, we compare the extracted answer by the \textsc{readers} against the answer provided by the analysts in our dataset (\textit{REQuestA}). 
We evaluate an extracted answer for correctness in three different modes. Let $a_{GT}$ denote the ground-truth answer to a question. In \textit{exact matching} mode, the extracted answer fully matches  $a_{GT}$. In \textit{partial matching} mode, the extracted answer partially matches (i.e., overlaps with) $a_{GT}$. In \textit{semantic matching} mode, the extracted answer has a cosine semantic similarity with  $a_{GT}$ that is greater than a predefined threshold. In our work, we apply a threshold of $0.5$~\cite{Ramage:09}. 
The first two modes evaluate correctness at a lexical level, whereas the last mode measures correctness based on meaning. 

In addition to reporting accuracy, we also report F1 measure -- 
another commonly-reported lexical metric in the QA literature~\cite{Cambazoglu:21}.
F1 is the harmonic mean computed as \hbox{$2*P*R/(P+R)$}, where $P$ is the precision and $R$ is the recall. 
We define $P$ as the number of overlapping tokens between the extracted answer and  $a_{GT}$ divided by the total number of tokens in the extracted answer. We define $R$ as the number of overlapping tokens between the extracted answer and  $a_{GT}$ divided by the total number of tokens in  $a_{GT}$. We report in EXPII overall F1-score averages for all questions.  

\vspace*{.2em}\sectopic{EXPIII. } This experiment answers \textbf{RQ3}. We report the execution of our approach with the most accurate models from the previous experiments. EXPIII is conducted on the Google Colaboratory cloud using the free plan with the following specifications: Intel(R) Xeon(R) CPU@2.20GHz, Tesla T4 GPU, and 13GB RAM.

%% file: subsections/RQs.tex
\subsection{Answers to the RQs}

\sectopic{RQ1. Which \textsc{retriever} has the highest accuracy in finding text that is most relevant to a given question?} 
RQ1 identifies the best-performing (i)~\textit{document} \textsc{retriever} and (ii)~\textit{passage} \textsc{retriever} to be applied in steps~1~and~3 of \textit{QAssist}, respectively. 
Tables~\ref{tab:rq1-a} and \ref{tab:rq1-b} \hbox{show the results of EXPI.}

\input{tables/RQ1-a}

In Table~\ref{tab:rq1-a}, traditional \textsc{retrievers} (TF-IDF and BM25) are clearly able to find the most relevant documents across all domains, thus achieving a perfect R@1. 
In comparison, our dense \textsc{retriever} (DistilBERT) has an average R@1 of 96.5\%, which is slightly worse than the traditional  \textsc{retrievers}. The reranking \textsc{retriever} achieves a perfect R@1 as well since it partially uses the results of BM25. 
In view of these results, we select BM25 as the \textsc{retriever} to use for step~1 of our approach, since BM25 is computationally more efficient than the reranking \textsc{retriever}. Compared to TF-IDF, BM25 is more robust~\cite{Whissell:11} and widely-applied in the QA literature~\cite{Thakur:21}.

In Table~\ref{tab:rq1-b}, we show the results for retrieving the most relevant $k$ text passages for $k=1, 3, 5, 10$. 
The upper part of the table provides the average results for our collection of six SRSs. The lower part of the table shows the results for retrieving passages from the most relevant external document. We recall from Section~\ref{sec:approach} that $\mathcal{T}_S$ denotes the set of passages within a given SRS and $\mathcal{T}_D$ denotes the passages in the most relevant external document from the corpus. In our dataset, an SRS has on average about 40 passages, whereas an external document has on average 53  passages.
Here, recall measures the presence of the relevant passage in the retrieved passages, whereas nDCG measures whether the relevant passage has a higher rank among the retrieved passages. In our analysis, we focus on recall, noting that rank does not play as significant a role for small values of $k$ ($\leq 3$) where our discussion of recall, below, leads us to.

We observe from Table~\ref{tab:rq1-b} that the reranking \textsc{retriever} outperforms the alternatives in the two metrics and for all $k$ values, except for the security domain as we elaborate later. 
We naturally see improvement in  recall with higher values of $k$. Concretely, the reranking \textsc{retriever} achieves for retrieving passages from the SRSs an average recall of 78.9\%, 90.1\%, 92.2\%, and 92.4\% at 
$k=1$, $k=3$, $k=5$, and $k=10$, respectively. The same \textsc{retriever} achieves for retrieving  passages from the external document an average recall of 77.0\%  at 
$k=1$, and 96.5\% at $k=3$, $k=5$, and $k=10$.

\input{tables/RQ1-b}

Selecting the best value of $k$ has practical implications. 
While higher $k$ values yield better recall, they entail additional effort for reviewing the results of \textit{QAssist}. For instance, selecting $k=10$ yields the best overall results, which implies that a stakeholder has more relevant context at their disposal for understanding and interpreting the requirements.
However, this comes at the cost of more time and effort needed to browse through the retrieved text passages. 
We deem $k=3$ as a reasonable compromise in our context, since the gain in recall at $k=5$ (in comparison to $k=3$) is merely $\approx$2 percentage points; selecting $k=5$ would imply browsing through two additional passages per question. That said, $k$ can be left as a user-configurable parameter, to be adjusted according to needs and the time budget available. 

The results show that the dense \textsc{retriever}, DistilBERT, performs on par with the reranking  \textsc{retriever} for the security domain. 
In our collection, the domain-specific corpus generated for security is the smallest among the corpora as it is generated from two SRSs, one of which is very small (SRS \#6). Furthermore, the number of passages analyzed in this domain is 23, compared to the aerospace and defence with an average of 42 and 94 passages, respectively. This observation suggests that the dense \textsc{retriever} is more effective when there is a fewer number of passages. 
The performance of the reranking \textsc{retriever} is in general better than that of the dense \textsc{retriever} for $k=3$. Consequently, we select the reranking \textsc{retriever} as the best-performing \hbox{alternative for step~3 of our approach.}
\pagebreak[4]

\begin{tcolorbox}[arc=0mm,width=\columnwidth,
                  top=1mm,left=1mm,  right=1mm, bottom=1mm,
                  boxrule=1pt] 
The answer to \textbf{RQ1} is that BM25 is the best document \textsc{retriever} with a perfect recall, and the reranking \textsc{retriever} is the best passage \textsc{retriever} with an average recall@$3$ of 90.1\% and 96.5\%  for SRSs and external (corpus) documents, respectively.
\end{tcolorbox}
\vspace*{.5em}

\sectopic{RQ2. Which \textsc{reader} produces the most accurate results for extracting the likely answer to a given question?} Table~\ref{tab:rq2} shows the results of EXPII, comparing the accuracy of the \textsc{readers} for extracting the answer to a given question.
Note that in RQ1, we focused on retrieving \emph{passages}, whereas in RQ2, we are interested in determining which \textsc{reader} identifies the most accurate \emph{text span} containing the answer within the passages already found.

The table shows that the most accurate \textsc{reader} varies depending on which matching mode we choose.
Considering the \textit{exact matching} mode, RoBERTa is the most accurate \textsc{reader}, followed by ALBERT, with an average accuracy of 24.6\%  
and 24.3\%, respectively. This finding is corroborated by the F1 measure. Nevertheless, both \textsc{readers} are outperformed by DistilBERT in the \textit{partial matching} mode which achieves the best average accuracy of 86.4\%. 

Noting their lexical nature, the exact and partial matching modes as well as the F1 measure have the drawback that they focus on whether the extracted answer is literally the same as the one in the ground truth rather than providing equivalent information~\cite{Risch:21}. 
For example, consider question Q1 in Fig.~\ref{fig:example}. The answer extracted for this question from the first passage of the domain-specific corpus (right side of the figure) could be the following: ``how much more massive the vehicle is with propellant than without''. This answer does not have a lexical overlap with the highlighted answer (shaded green in the figure), despite considerable similarity in meaning. For such cases, lexical metrics would evaluate the extracted answer as incorrect. To better assess the performance of the \textsc{readers} in our context, where users may be seeking all closely relevant information, we further report results for the \textit{semantic matching} mode. 
Using the \textit{semantic matching} mode would lead us to the same conclusion as that offered by \textit{exact matching} and F1. That is, ALBERT and RoBERTa have the highest average accuracy of 84.2\% and 84.0\%, respectively. 
Despite the similar behavior of the two models, ALBERT considerably outperforms RoBERTa in \textit{partial matching} mode with an average percentage points of $\approx$19\%. 
We thus select ALBERT as the \hbox{best-performing \textsc{reader} for answer extraction.}

{\color{black}
Since Wikipedia has been used for pre-training BERT and many variants thereof, and considering that part of our question-answer pairs originate from Wikipedia, we show that answer extraction in our approach is still accurate for content that originates from sources different from Wikipedia. Recall from  Table~\ref{tab:docCollection} that REQuestA contains a total of 189 (= 86 + 103) question-answer pairs from SRSs and another 198 (= 87 + 111) pairs from Wikipedia articles. The 189 question-answer pairs from the SRSs are independent from Wikipedia. The performance of BERT-based models over these pairs is a representative indicator for non-Wikipedia content. 

In Table~\ref{tab:rq2}, we further provide a breakdown of the \textsc{reader} results based on the origin of the question-answer pairs. We denote SRS-based questions as $q_{S}$ and domain-based questions (which, in our case study, are sourced from Wikipedia) as $q_{D}$. The table shows that all models achieve on-par or better accuracy over $q_{S}$ compared to $q_{D}$. Based on the breakdown in Table~\ref{tab:rq2}, we conclude that the exposure of BERT-based models to Wikipedia during pre-training is unlikely to have influenced our performance results.

}
\begin{tcolorbox}[arc=0mm,width=\columnwidth,
                  top=1mm,left=1mm,  right=1mm, bottom=1mm,
                  boxrule=1pt] 
The answer to \textbf{RQ2} is that considering both lexical and semantic measures, ALBERT provides the best overall trade-off for answer extraction with an average accuracy of $\approx$24\% in the \textit{exact matching} mode, $\approx$79\% in the \textit{partial matching} mode, and $\approx$84\% in the \textit{semantic matching} mode.  
\end{tcolorbox}

\input{tables/RQ2}

\sectopic{RQ3. Does \textit{QAssist} run in practical time?} 
To answer RQ3, we discuss the execution time of our approach based on the conclusions from RQ1 and RQ2 and the setup described under \textit{EXPIII} in Section~\ref{subsec:evaluationProc}. Based on RQ1, we select BM25 as the document \textsc{retriever} and the reranking method as the passage \textsc{retriever}. For answer extraction, based on RQ2, we select ALBERT as the \textsc{reader}. 
With these choices, we report the execution time for each step of \textit{QAssist} (Fig.~\ref{fig:approach}).

Retrieving the most relevant document from the corpora created for the aerospace, defence, and security domains (step~1) requires 2.06, 1.37, and 0.08 seconds, respectively. The time required  in step~2 for splitting a document into tokens and sentences is comparatively negligible. For retrieving relevant passages in step~3, we note that the six SRSs in our study vary in size from small (SRS\#6 with 32 requirements) to large (SRS\#2 with 1041 requirements). Similarly, the Wikipedia articles (making up the domain-specific corpora) from which we retrieve passages vary in size, as shown previously in Table~\ref{tab:docCollection}. 
For our dataset, the time required for retrieving passages from an SRS is 2.27 seconds for the smallest SRS and 6.43 seconds for the largest. For corpus articles, the average time for passage retrieval is 2.62 seconds. Extracting  answers from passages, i.e., step~4, takes an average of 1.1 seconds.

In addition to the above-reported execution times, there is a one-time loading overhead for the \textsc{reader}, as shown in the last column of Table~\ref{tab:rq2}.
For ALBERT (best \textsc{reader} from RQ2), this overhead is $\approx$3.2 minutes. We deem this overhead acceptable considering that, once the \textsc{reader} has been loaded, the user can ask as many questions as desired.

Excluding the overhead for loading the \textsc{reader}, answering an individual question, when averaged across all questions in our dataset, takes 10.36 seconds. We believe this execution time is reasonable for most practical situations. Moreover, the execution time can be improved if one has access to more powerful computing resources than ours (Google Colab's free plan, as noted in Section~\ref{subsec:evaluationProc}).

\vspace*{.2em}
\begin{tcolorbox}[arc=0mm,width=\columnwidth,
                  top=1mm,left=1mm,  right=1mm, bottom=1mm,
                  boxrule=1pt] 
When run on Google Colab's free plan, our approach takes an average of 10.36 seconds to answer an individual question. In addition, one has to provision for a one-time overhead of 3.2 minutes to load the required language model (ALBERT). We find this level of performance practical for question answering over requirements. Performance can be further improved with more powerful computational resources for language models.
\end{tcolorbox}
\vspace*{.2em}

\section{Comparison with Broad-based Search Engines} \label{sec:google}

{\color{black}An intuitive way for QA during the analysis of an SRS would be to pose the questions to a (broad-based) search engine such as Google.  In the context of our work, search engines are generally not very effective for two main reasons. First, answers to domain-specific questions can reside in company-specific documents which are unlikely to be accessible to search engines. 
Our approach, in contrast, gives analysts the possibility to plug company-specific documents into the QA system. 
Second, the lack of domain-specificity in search engines can easily result in misleading answers. For example, an online search for ``rocket mass'' instead of ``wet mass'' to answer Q1 in Fig.~\ref{fig:example} would point the analyst to the design of a rocket mass heater\footnote{\url{https://en.wikipedia.org/wiki/Rocket\_mass\_heater}}, which is not relevant to the space domain. 
Unlike search engines, our approach is scoped to the original SRS and any external knowledge resources selected by the user. As such, questions are implicitly disambiguated as long as the external knowledge resources are domain-specific. To further illustrate, consider the question ``What is NEAT?''.  Posing this question online would lead to irrelevant results due to the ambiguous abbreviation, whereas posing the same question to our approach would retrieve the definition of ``Near-Earth Asteroid Tracking'' -- inline with the SRS content.

To empirically assess the success rate of search engines in our problem context, we posed to Google from our dataset a total of 50 verbatim questions. Of these, 20 questions were SRS-based and 30 were domain-based. The authors independently investigated whether the top-3 retrieved documents by Google contained the correct answer as per our ground truth. Out of the 50 questions, we found that 16 questions were answered correctly by Google, leading to a success rate of 32\%. From the 16 correctly answered questions, 14 were domain-based. We note that the domain-based questions in our dataset, REQuestA, originate from Wikipedia articles, which search engines have access to and can crawl. The outcome would most likely have been different had the external knowledge resource not been public. Therefore, in addition to the need for explicit disambiguation as discussed above, the success rate of search engines is likely to be affected by the public accessibility of the documents that should be considered during QA. 
%
In conclusion, we believe that search engines are currently not the best alternative for QA over specialized and proprietary material -- a situation that is common in RE.}

%% file: tables/RQ1-a.tex
\begin{table}
\centering
\caption{R@$1$ of \textit{Document} \textsc{Retriever} (\textbf{RQ1}).}
\label{tab:rq1-a}
\begin{threeparttable}[t]
\begin{tabularx}{0.48\textwidth}{@{} p{0.08\textwidth} @{\hskip 0.5em} p{0.05\textwidth} @{\hskip 0.5em} *{4}{>{\centering\arraybackslash}X}@{}}
    \toprule
Domain &  $|\mathcal{D}|^\dag$ & TF-IDF & BM25 & Dense & Reranking \\ 
\midrule
Aerospace & 1158 & \textbf{100}  & \textbf{100} & 99.0  & \textbf{100}   \\   
Defence & 781 & \textbf{100}  & \textbf{100} & 98.9  & \textbf{100} \\
Security & 50  & \textbf{100} & \textbf{100}   & 91.7 & \textbf{100}\\ 
\bottomrule
\end{tabularx}
\begin{tablenotes}
     \item[$\dag$] $|\mathcal{D}|$ is the number of articles in the corpus ($\mathcal{D}$) of Wikipedia articles. 
     \end{tablenotes}
 \end{threeparttable}
 \vspace*{-1.2em}
\end{table}

%% file: tables/RQ1-b.tex
\begin{table}
\centering
\caption{Accuracy of \textit{Passage} \textsc{Retriever} 
(\textbf{RQ1}).} 
\label{tab:rq1-b}
\begin{threeparttable}[t]
\begin{tabularx}{0.5\textwidth}{@{}l*{10}{>{\centering\arraybackslash}X}@{}}
\toprule
&& \multicolumn{2}{c}{Top-1} & \multicolumn{2}{c}{Top-3} & \multicolumn{2}{c}{Top-5} & \multicolumn{2}{c}{Top-10} \\ \cmidrule(lr){3-4}\cmidrule(lr){5-6}\cmidrule(lr){7-8}\cmidrule(lr){9-10}
\multicolumn{2}{c}{From $\mathcal{T}_S$} 
& R & nDCG & R  & nDCG & R &  nDCG & R &  nDCG\\
\midrule
\multicolumn{2}{c}{(1)}  & 60.3 & 60.3 
      & 78.6 & 70.9 
      & 84.3 & 73.3 
      & 89.8 & 75.1 \\

\multicolumn{2}{c}{(2)} & 62.8  & 62.8 
     & 78.5  & 71.9 
     & 85.4  & 74.6 
     & \textbf{92.4} & 76.9\\

\multicolumn{2}{c}{(3)} & 52.9 & 52.9 
     & 81.2  & 70.3 
     & 86.5  & 72.5 
     & 88.5 & 73.2\\

\multicolumn{2}{c}{(4)} & \textbf{78.9} & \textbf{78.9}
     & \textbf{90.1}  & \textbf{85.8} 
     & \textbf{92.2}  & \textbf{86.6} 
     & \textbf{92.4} & \textbf{86.7}\\
\midrule
\midrule
\multicolumn{2}{c}{From $\mathcal{T}_D$} 
& R  & nDCG& R  &  nDCG& R  & nDCG & R &  nDCG \\
\midrule
\multirow{4}{*}{\rotatebox[origin=c]{90}{Aerospace}}& (1) & 43.6 & 43.6 
& 64.7 & 56.0 
& 68.6 & 57.5 
& 91.5 & \textbf{94.9} \\

&(2) & 50.5  & 50.5 
     & 83.0 &  70.1 
     & 91.7 & 73.6 
     & \textbf{95.0} & 74.7 \\

&(3) & 66.7 &  66.7 
     & 87.3 &  79.5 
     & 90.6 & 80.8 
     & 91.3 &  81.0 \\

&(4)& \textbf{75.1} & \textbf{75.1} 
    & \textbf{95.0}  & \textbf{87.3} 
    & \textbf{95.0} &  \textbf{87.3} 
    & \textbf{95.0}  & 87.3 \\
\midrule

\multirow{4}{*}{\rotatebox[origin=c]{90}{Defence}}&(1)  & 41.9 & 41.9 
& 62.1 & 54.1 
& 66.7 & 55.9 
& 86.3  & 62.2\\

&(2)  & 38.0  & 38.0 
    & 81.2 & 64.1 
    & 89.3 &  67.3 
    & \textbf{94.6} & 69.1\\
    
&(3)  & \textbf{77.2}  &  \textbf{77.2} 
      & 89.8  & 84.7 
      & \textbf{91.2}  & 85.2
      & 92.5 &  85.6\\

&(4) & 76.0  & 76.0 
    & \textbf{94.6}  & \textbf{87.6} 
    & 94.6  & \textbf{87.6}
    & \textbf{94.6} &  \textbf{87.6} \\
    
    \midrule 
  \multirow{4}{*}{\rotatebox[origin=c]{90}{Security}}&(1) 
  & 33.4  & 33.4 
  & 70.0& 53.8 
  & 70.0 &  53.8 
  & 80.0 & 57.4\\
  
&(2)  & 43.3  &  43.3 
& 70.0 & 59.3
& \textbf{100}  & 71.3 
& \textbf{100} & 71.3\\

&(3) & 63.3  &  63.3 
& \textbf{100}  &  85.2 
& \textbf{100}  &  85.2 
& \textbf{100} & 85.2\\

&(4)& \textbf{80.0}  & \textbf{80.0} & 
\textbf{100}&  \textbf{92.6} 
& \textbf{100}&  \textbf{92.6} 
& \textbf{100} & \textbf{92.6}\\
\bottomrule
\end{tabularx}
\begin{tablenotes}
     \item (1) TF-IDF, (2) BM25, (3) Dense, and (4) Reranking. 
     \end{tablenotes}
 \end{threeparttable}
 \vspace*{-2em}
 \end{table}

%% file: tables/RQ2.tex
\begin{table}
\centering
\caption{\textsc{Reader} Accuracy Results (\textbf{RQ2});  
table further shows loading time for \textsc{readers} (a consideration for \textbf{RQ3}). }
\label{tab:rq2}
\begin{threeparttable}[t]
\begin{tabularx}{0.48\textwidth} {@{}l*{5}{>{\centering\arraybackslash}X}@{}}
\toprule
\multirow{2}{*}{Model} & \multicolumn{3}{c}{Accuracy} & \multirow{2}{*}{F1} & \multirow{2}{*}{Time} \\ 
\cmidrule(l){2-4}
& \textit{Exact} & \textit{Partial} & \textit{Semantic} & & \\ 
\toprule
ALBERT & 24.3 & 79.1 & \textbf{84.2} & 64.6& \multirow{3}{*}{193.2} \\
\cmidrule(l){1-1}
\multicolumn{1}{r}{$q_{S}$} &31.7&78.0&86.4&67.6\\
\multicolumn{1}{r}{$q_{D}$} &17.2 &80.2 & 82.1 &
61.7 \\
\midrule 
BERT & 21.4 & 70.6 & 82.9 & 63.1& \multirow{3}{*}{32.2} \\ 
\cmidrule(l){1-1}
\multicolumn{1}{r}{$q_{S}$} &28.3&70.4&83.8&
66.4\\
\multicolumn{1}{r}{$q_{D}$} &14.8&70.8&82.1&59.9 \\
\midrule
DistilBERT & 23.0 & \textbf{86.4} & 75.9 & 61.0& \multirow{3}{*}{5.8} \\
\cmidrule(l){1-1}
\multicolumn{1}{r}{$q_{S}$} &32.7&86.4&77.3&67.5\\
\multicolumn{1}{r}{$q_{D}$} &13.7&86.5&74.6&54.8 \\
\midrule
ELECTRA& 21.1 & 81.3 & 81.0 & 60.1& \multirow{3}{*}{19.1} \\ 
\cmidrule(l){1-1}
\multicolumn{1}{r}{$q_{S}$} &31.2&82.1&80.6&65.0 \\
\multicolumn{1}{r}{$q_{D}$} &11.5&80.5&81.4&55.4 \\
\midrule
MiniLM & 23.3 & 73.3 & 82.4 & 63.4& \multirow{3}{*}{5.0 }\\
\cmidrule(l){1-1}
\multicolumn{1}{r}{$q_{S}$} &32.4&73.6&82.6&66.1 \\
\multicolumn{1}{r}{$q_{D}$} &14.6&73.0&82.2&60.9 \\
\midrule
RoBERTa& \textbf{24.6} & 60.2 & 84.0 & \textbf{65.2} & \multirow{3}{*}{11.6} \\ 
\cmidrule(l){1-1}
\multicolumn{1}{r}{$q_{S}$} &32.8&61.0&84.3&70.1 \\
\multicolumn{1}{r}{$q_{D}$} &16.8&59.4&83.7&60.5 \\
\bottomrule
\end{tabularx}
\begin{tablenotes}
     \item[] \textcolor{black}{\it The table reports performance results for all question-answer pairs as well as for SRS-based ($q_{S}$) and domain-based ($q_{D}$) pairs separately.}
     \end{tablenotes}
 \end{threeparttable}
\vspace*{-1em}
\end{table}

%% file: sections/threats.tex
\section{Threats to Validity}\label{sec:threats}
The validity concerns most pertinent to our evaluation are internal
and external validity.

\sectopic{Internal Validity.} The main concern regarding internal validity is dataset bias. To mitigate bias, the authors ensured that they were  not involved in dataset construction; this task was done exclusively by third parties (non-authors) who had no exposure to our technical solution. 



\sectopic{External Validity.} Our evaluation is based on a dataset containing six industrial SRSs and spanning three different application domains. The results we obtained across these SRSs and domains combined with the comparatively large size of our QA dataset 
provide confidence about the generalizability of our empirical findings.
Additional experimentation is nevertheless important to further mitigate  external-validity threats.  

%% file: sections/related.tex
\section{Related Work}\label{sec:related}

In this section, we position our work in the existing literature on QA as studied by the RE and NLP communities. 

\sectopic{QA in RE. } 
There has been only limited research where QA is applied for addressing RE problems. Existing works focus on requirements traceability~\cite{Mader:13,Pruski:15,Lin:17}, identifying compliance requirements~\cite{Sleimi:19,Abualhaija:22}, 
and extracting information from online forums~\cite{Kanchev:17}. These techniques are mostly IR-based, with the exception of \cite{Abualhaija:22}, which, like our approach, uses machine reading comprehension (MRC).
Our approach differs from \cite{Abualhaija:22} both in its purpose and also in how it employs MRC. First, whereas \cite{Abualhaija:22} focuses on QA over legal provisions (e.g., privacy regulations), our approach deals with QA over SRSs. 
Second, \cite{Abualhaija:22} is limited in that it applies MRC to a-priori-specified documents only. Our approach can, in contrast, mine domain-related content from Wikipedia in an attempt to make tacit domain knowledge explicit and thereby handle questions that would go unanswered if the scope of search for answers was limited to the SRS under analysis only.

In terms of QA datasets, not many such datasets are available in RE. Abualhaija et al.'s dataset of 107 question-and-answer pairs~\cite{Abualhaija:22} is built over legal documents. 
In contrast, our dataset, \textit{REQuestA}, is built over SRSs. To our knowledge, \textit{REQestA} is the first dataset of its kind, providing a total of 387 question-and-answer pairs on industrial requirements.

Malviya et al.~\cite{Malviya:17} investigate questions that requirements engineers typically ask throughout the development process. They collect through a survey with industry practitioners a set of 159 questions, grouped into nine different categories such as project management and quality assessment. 
Malviya~et~al.'s questions are broad and can crosscut several artifacts in the development life cycle. Our work focuses specifically on clarification questions asked  over SRSs and associated domain-knowledge resources; our objective here is developing automated QA technologies that can answer such questions.

\sectopic{QA in NLP. } QA tasks in the NLP literature include question classification, answer extraction, and question generation~\cite{Hao:22,Yusuf:22,Jin:19,Diefenbach:20}. 
Answer extraction is considered to be the main QA task in NLP~\cite{Ojokoh:18}.
Recent advances in QA answer extraction include fine-tuning large scale 
language models such as BERT, RoBERTa, and ALBERT~\cite{Jing:19,Wulamu:19,Ren:20,Parshakova:19}. 
Inspired by the NLP literature, we apply in our work the QA models reported in a recent QA benchmark~\cite{Thakur:21}.
Several existing QA datasets curated from generic text 
are publicly available. These datasets  include SQuAD~\cite{SQUAD:16}, GLUE~\cite{Wang:18}, 
and TriviaQA~\cite{Joshi:17}. 
There are also some domain-specific datasets, 
e.g., for the medical~\cite{He:19} and railway~\cite{Hu:20} domains. 
For the same reasons mentioned earlier when discussing related work in RE, none of the available datasets in NLP are suitable for our needs in this paper. 

Language models have been employed for various text generation tasks~\cite{Pan:19}, including question generation (QG)~\cite{Kumar:19,Raffel:19}. 
QG models have enabled researchers in many fields to automatically generate their own synthetic QA datasets~\cite{Liu:21,Bartolo:21,Lelkes:21,Gupta:22}. Our dataset was partially generated using QG. To our knowledge, QG has not been attempted in RE before.

Our work is distinguished from QA research in NLP in that we provide an end-to-end solution. Our approach covers all QA steps starting from posing a question down to providing the most relevant passages and potential answers. Foundational research in NLP often focuses on individual QA steps, e.g., IR-based text retrieval  or MRC-based answer extraction. Our work does not contribute to the foundations for QA. Nevertheless, our motivating use case (QA over requirements), our combination of NLP technologies, the flexibility to build domain-specific corpora and consult them during QA, and our extensive empirical evaluation of QA in an RE context are, to the best of our knowledge, new.

%% file: sections/conclusion.tex
\section{Conclusion}\label{sec:conclusion}

In this paper, we proposed \textit{QAssist} -- an AI-based question-answering (QA) system to support the analysis of natural-language requirements. 
Given a question, \textit{QAssist} retrieves relevant text passages from both the requirements document 
being analyzed as well as an external source of domain knowledge. \textit{QAssist} further highlights the likely answer to the question in each retrieved text passage. The flexibility to incorporate an external knowledge source into the QA process enables \textit{QAssist} to answer otherwise unanswerable questions related to the tacit domain information assumed by the requirements. When a domain-knowledge resource is absent, \textit{QAssist} automatically builds one by mining Wikipedia articles, using the terminology in the requirements being analyzed to guide the mining process.
%
To evaluate \textit{QAssist}, we created through third-party annotators a QA dataset, named \textit{REQuestA}. Both \textit{QAssist} and \textit{REQuestA} are publicly available~\cite{qassist-rep}.
%
%
Our empirical results indicate that
\textit{QAssist} localizes the answer to a posed question to three passages within the requirements document and within the external domain-knowledge resource with an average recall of 90.1\% and 96.5\%, respectively.  Narrowing the scope to these passages, \textit{QAssist} has an average accuracy of 84.2\% in pinpointing the actual answer.

In future work, we would like to conduct user studies to better understand how practitioners would interact with requirements documents when equipped with a QA tool. 
Another future direction is to experiment with emerging QA methods in NLP that are capable of producing a ``no answer'' outcome when a question is not answerable \hbox{with sufficient accuracy.}

%% file: sections/ack.tex
\textcolor{black}{
\sectopic{Acknowledgements.} This work was funded by Luxembourg's National Research Fund (FNR) under 
the grant BRIDGES18/IS/12632261 
and NSERC of Canada under the Discovery and Discovery Accelerator programs. We are grateful to the research and development team at QRA Corp. for valuable insights and assistance.
}

%% file: paper.bbl
\begin{thebibliography}{10}
\providecommand{\url}[1]{#1}
\csname url@samestyle\endcsname
\providecommand{\newblock}{\relax}
\providecommand{\bibinfo}[2]{#2}
\providecommand{\BIBentrySTDinterwordspacing}{\spaceskip=0pt\relax}
\providecommand{\BIBentryALTinterwordstretchfactor}{4}
\providecommand{\BIBentryALTinterwordspacing}{\spaceskip=\fontdimen2\font plus
\BIBentryALTinterwordstretchfactor\fontdimen3\font minus
  \fontdimen4\font\relax}
\providecommand{\BIBforeignlanguage}[2]{{%
\expandafter\ifx\csname l@#1\endcsname\relax
\typeout{** WARNING: IEEEtran.bst: No hyphenation pattern has been}%
\typeout{** loaded for the language `#1'. Using the pattern for}%
\typeout{** the default language instead.}%
\else
\language=\csname l@#1\endcsname
\fi
#2}}
\providecommand{\BIBdecl}{\relax}
\BIBdecl

\bibitem{vanLamsweerde:09}
A.~van Lamsweerde, \emph{Requirements Engineering: {From} System Goals to {UML}
  Models to Software Specifications}, 1st~ed.\hskip 1em plus 0.5em minus
  0.4em\relax Wiley, 2009.

\bibitem{Pohl:10}
K.~Pohl, \emph{Requirements Engineering}, 1st~ed.\hskip 1em plus 0.5em minus
  0.4em\relax Springer, 2010.

\bibitem{Zhao:20}
L.~Zhao, W.~Alhoshan, A.~Ferrari, K.~J. Letsholo, M.~A. Ajagbe, E.-V. Chioasca,
  and R.~T. Batista-Navarro, ``Natural language processing (nlp) for
  requirements engineering: A systematic mapping study,'' \emph{arXiv preprint
  arXiv:2004.01099}, 2020.

\bibitem{Ferrari:19}
A.~Ferrari and A.~Esuli, ``An {NLP} approach for cross-domain ambiguity
  detection in requirements engineering,'' \emph{Automated Software
  Engineering}, vol.~26, no.~3, 2019.

\bibitem{Ezzini:21}
S.~Ezzini, S.~Abualhaija, C.~Arora, M.~Sabetzadeh, and L.~C. Briand, ``Using
  domain-specific corpora for improved handling of ambiguity in requirements,''
  in \emph{2021 IEEE/ACM 43rd International Conference on Software
  Engineering}, 2021.

\bibitem{Dalpiaz:18}
F.~Dalpiaz, I.~Schalk, and G.~Lucassen, ``Pinpointing ambiguity and
  incompleteness in requirements engineering via information visualization and
  {NLP},'' in \emph{Proceedings of the 24th Working Conference on Requirements
  Engineering: Foundation for Software Quality}, 2018.

\bibitem{Arora:19}
C.~Arora, M.~Sabetzadeh, and L.~C. Briand, ``An empirical study on the
  potential usefulness of domain models for completeness checking of
  requirements,'' \emph{Empirical Software Engineering}, vol.~24, no.~4, pp.
  2509--2539, 2019.

\bibitem{Hadar:19}
I.~Hadar, A.~Zamansky, and D.~M. Berry, ``The inconsistency between theory and
  practice in managing inconsistency in requirements engineering,''
  \emph{Empirical Software Engineering}, vol.~24, no.~6, pp. 3972--4005, 2019.

\bibitem{Jurafsky:20}
D.~Jurafsky and J.~H. Martin, \emph{Speech and Language Processing}, 3rd~ed.,
  2020, \url{https://web.stanford.edu/~jurafsky/slp3/}(visited 2021-06-04).

\bibitem{Arora:17}
C.~Arora, M.~Sabetzadeh, L.~Briand, and F.~Zimmer, ``Automated extraction and
  clustering of requirements glossary terms,'' \emph{IEEE Transactions on
  Software Engineering}, vol.~43, no.~10, 2017.

\bibitem{Ezzini2022wikidominer}
S.~Ezzini, S.~Abualhaija, and M.~Sabetzadeh, ``Wikidominer: Wikipedia
  domain-specific miner,'' in \emph{Proceedings of the 17th joint meeting of
  the European Software Engineering Conference and the ACM SIGSOFT Symposium on
  the Foundations of Software Engineering}, 2022.

\bibitem{Zhu:21}
F.~Zhu, W.~Lei, C.~Wang, J.~Zheng, S.~Poria, and T.-S. Chua, ``Retrieving and
  reading: A comprehensive survey on open-domain question answering,''
  \emph{arXiv preprint arXiv:2101.00774}, 2021.

\bibitem{qassist-rep}
\BIBentryALTinterwordspacing
``Replication package,'' 2022. [Online]. Available:
  \url{https://gitlab.uni.lu/sezzini/QAssist/}
\BIBentrySTDinterwordspacing

\bibitem{Maletic:09}
J.~I. Maletic and M.~L. Collard, ``Tql: A query language to support
  traceability,'' in \emph{2009 ICSE workshop on traceability in emerging forms
  of software engineering}.\hskip 1em plus 0.5em minus 0.4em\relax IEEE, 2009,
  pp. 16--20.

\bibitem{Mader:13}
P.~M{\"a}der and J.~Cleland-Huang, ``A visual language for modeling and
  executing traceability queries,'' \emph{Software \& Systems Modeling},
  vol.~12, no.~3, pp. 537--553, 2013.

\bibitem{Pruski:15}
P.~Pruski, S.~Lohar, W.~Goss, A.~Rasin, and J.~Cleland-Huang, ``Tiqi: answering
  unstructured natural language trace queries,'' \emph{Requirements
  Engineering}, vol.~20, no.~3, pp. 215--232, 2015.

\bibitem{Lin:17}
J.~Lin, Y.~Liu, J.~Guo, J.~Cleland-Huang, W.~Goss, W.~Liu, S.~Lohar,
  N.~Monaikul, and A.~Rasin, ``Tiqi: A natural language interface for querying
  software project data,'' in \emph{2017 32nd IEEE/ACM International Conference
  on Automated Software Engineering}.\hskip 1em plus 0.5em minus 0.4em\relax
  IEEE, 2017, pp. 973--977.

\bibitem{Malviya:17}
S.~Malviya, M.~Vierhauser, J.~Cleland-Huang, and S.~Ghaisas, ``What questions
  do requirements engineers ask?'' in \emph{2017 IEEE 25th International
  Requirements Engineering Conference}.\hskip 1em plus 0.5em minus 0.4em\relax
  IEEE, 2017, pp. 100--109.

\bibitem{Abualhaija:22}
S.~Abualhaija, C.~Arora, A.~Sleimi, and L.~Briand, ``Automated question
  answering for improved understanding of compliance requirements: A
  multi-document study,'' in \emph{In Proceedings of the 30th IEEE
  International Requirements Engineering Conference, Melbourne, Australia 15-19
  August 2022}, 2022.

\bibitem{Soares:20}
M.~A.~C. Soares and F.~S. Parreiras, ``A literature review on question
  answering techniques, paradigms and systems,'' \emph{Journal of King Saud
  University-Computer and Information Sciences}, vol.~32, no.~6, pp. 635--646,
  2020.

\bibitem{SQUAD:16}
P.~Rajpurkar, J.~Zhang, K.~Lopyrev, and P.~Liang, ``Squad: 100,000+ questions
  for machine comprehension of text,'' \emph{arXiv preprint arXiv:1606.05250},
  2016.

\bibitem{Joshi:17}
M.~Joshi, E.~Choi, D.~S. Weld, and L.~Zettlemoyer, ``Triviaqa: A large scale
  distantly supervised challenge dataset for reading comprehension,''
  \emph{arXiv preprint arXiv:1705.03551}, 2017.

\bibitem{Kwiatkowski:19}
T.~Kwiatkowski, J.~Palomaki, O.~Redfield, M.~Collins, A.~Parikh, C.~Alberti,
  D.~Epstein, I.~Polosukhin, J.~Devlin, K.~Lee \emph{et~al.}, ``Natural
  questions: a benchmark for question answering research,'' \emph{Transactions
  of the Association for Computational Linguistics}, vol.~7, pp. 453--466,
  2019.

\bibitem{Pampari:18}
A.~Pampari, P.~Raghavan, J.~Liang, and J.~Peng, ``emrqa: A large corpus for
  question answering on electronic medical records,'' in \emph{Proceedings of
  the 2018 Conference on Empirical Methods in Natural Language Processing},
  2018, pp. 2357--2368.

\bibitem{He:19}
J.~He, M.~Fu, and M.~Tu, ``Applying deep matching networks to chinese medical
  question answering: a study and a dataset,'' \emph{BMC medical informatics
  and decision making}, vol.~19, no.~2, pp. 91--100, 2019.

\bibitem{Tian:19}
Y.~Tian, W.~Ma, F.~Xia, and Y.~Song, ``Chimed: A chinese medical corpus for
  question answering,'' in \emph{Proceedings of the 18th BioNLP Workshop and
  Shared Task}, 2019, pp. 250--260.

\bibitem{Hu:20}
Z.~Hu, ``Research and implementation of railway technical specification
  question answering system based on deep learning,'' in \emph{2020 IEEE 5th
  Information Technology and Mechatronics Engineering Conference (ITOEC)},
  2020, pp. 5--9.

\bibitem{Chen:17}
D.~Chen, A.~Fisch, J.~Weston, and A.~Bordes, ``Reading {W}ikipedia to answer
  open-domain questions,'' in \emph{Proceedings of the 55th Annual Meeting of
  the Association for Computational Linguistics}.\hskip 1em plus 0.5em minus
  0.4em\relax Association for Computational Linguistics, 2017, pp. 1870--1879.

\bibitem{McGill:83}
M.~McGill and G.~Salton, \emph{Introduction to Modern Information
  Retrieval}.\hskip 1em plus 0.5em minus 0.4em\relax McGraw-Hill, 1983.

\bibitem{Devlin:18}
J.~Devlin, M.-W. Chang, K.~Lee, and K.~Toutanova, ``{BERT}: Pre-training of
  deep bidirectional transformers for language understanding,'' 2018.

\bibitem{Liu:19}
S.~Liu, X.~Zhang, S.~Zhang, H.~Wang, and W.~Zhang, ``Neural machine reading
  comprehension: Methods and trends,'' \emph{Applied Sciences}, vol.~9, no.~18,
  p. 3698, 2019.

\bibitem{Manning:08}
C.~Manning, P.~Raghavan, and H.~Schutze, \emph{Introduction to Information
  Retrieval}, 1st~ed.\hskip 1em plus 0.5em minus 0.4em\relax Cambridge
  University Press, 2008.

\bibitem{Jones:72}
K.~S. Jones, ``A statistical interpretation of term specificity and its
  application in retrieval,'' \emph{Journal of documentation}, 1972.

\bibitem{Robertson:09}
S.~Robertson and H.~Zaragoza, \emph{The probabilistic relevance framework: BM25
  and beyond}.\hskip 1em plus 0.5em minus 0.4em\relax Now Publishers Inc, 2009.

\bibitem{Thakur:21}
N.~Thakur, N.~Reimers, A.~R{\"u}ckl{\'e}, A.~Srivastava, and I.~Gurevych,
  ``Beir: A heterogenous benchmark for zero-shot evaluation of information
  retrieval models,'' \emph{arXiv preprint arXiv:2104.08663}, 2021.

\bibitem{Nogueira:19}
R.~Nogueira and K.~Cho, ``Passage re-ranking with bert,'' \emph{arXiv preprint
  arXiv:1901.04085}, 2019.

\bibitem{Wang:21}
K.~Wang, N.~Thakur, N.~Reimers, and I.~Gurevych, ``Gpl: Generative pseudo
  labeling for unsupervised domain adaptation of dense retrieval,'' \emph{arXiv
  preprint arXiv:2112.07577}, 2021.

\bibitem{Zhuang:21}
S.~Zhuang and G.~Zuccon, ``Dealing with typos for bert-based passage retrieval
  and ranking,'' in \emph{Proceedings of the 2021 Conference on Empirical
  Methods in Natural Language Processing}, 2021, pp. 2836--2842.

\bibitem{chen:20}
D.~Chen and W.-t. Yih, ``Open-domain question answering,'' in \emph{Proceedings
  of the 58th Annual Meeting of the Association for Computational Linguistics:
  Tutorial Abstracts}.\hskip 1em plus 0.5em minus 0.4em\relax Online:
  Association for Computational Linguistics, 2020, pp. 34--37.

\bibitem{Lewis:20}
P.~Lewis, E.~Perez, A.~Piktus, F.~Petroni, V.~Karpukhin, N.~Goyal,
  H.~K{\"u}ttler, M.~Lewis, W.-t. Yih, T.~Rockt{\"a}schel \emph{et~al.},
  ``Retrieval-augmented generation for knowledge-intensive nlp tasks,''
  \emph{Advances in Neural Information Processing Systems}, vol.~33, pp.
  9459--9474, 2020.

\bibitem{Pan:09}
S.~J. Pan and Q.~Yang, ``A survey on transfer learning,'' \emph{IEEE
  Transactions on knowledge and data engineering}, vol.~22, no.~10, pp.
  1345--1359, 2009.

\bibitem{Petroni:19}
F.~Petroni, T.~Rockt{\"a}schel, P.~Lewis, A.~Bakhtin, Y.~Wu, A.~H. Miller, and
  S.~Riedel, ``Language models as knowledge bases?'' \emph{arXiv preprint
  arXiv:1909.01066}, 2019.

\bibitem{Vaswani:17}
A.~Vaswani, N.~Shazeer, N.~Parmar, J.~Uszkoreit, L.~Jones, A.~N. Gomez,
  L.~Kaiser, and I.~Polosukhin, ``Attention is all you need,'' \emph{arXiv
  preprint arXiv:1706.03762}, 2017.

\bibitem{Clark:20}
K.~Clark, M.-T. Luong, Q.~V. Le, and C.~D. Manning, ``Electra: Pre-training
  text encoders as discriminators rather than generators,'' \emph{arXiv
  preprint arXiv:2003.10555}, 2020.

\bibitem{Lan:19}
Z.~Lan, M.~Chen, S.~Goodman, K.~Gimpel, P.~Sharma, and R.~Soricut, ``Albert: A
  lite bert for self-supervised learning of language representations,''
  \emph{arXiv preprint arXiv:1909.11942}, 2019.

\bibitem{Sanh:19}
V.~Sanh, L.~Debut, J.~Chaumond, and T.~Wolf, ``Distilbert, a distilled version
  of bert: smaller, faster, cheaper and lighter,'' \emph{arXiv preprint
  arXiv:1910.01108}, 2019.

\bibitem{Wang:20a}
W.~Wang, F.~Wei, L.~Dong, H.~Bao, N.~Yang, and M.~Zhou, ``Minilm: Deep
  self-attention distillation for task-agnostic compression of pre-trained
  transformers,'' \emph{Advances in Neural Information Processing Systems},
  vol.~33, pp. 5776--5788, 2020.

\bibitem{Liu:19a}
Y.~Liu, M.~Ott, N.~Goyal, J.~Du, M.~Joshi, D.~Chen, O.~Levy, M.~Lewis,
  L.~Zettlemoyer, and V.~Stoyanov, ``Roberta: A robustly optimized bert
  pretraining approach,'' \emph{arXiv preprint arXiv:1907.11692}, 2019.

\bibitem{Gou:21}
J.~Gou, B.~Yu, S.~J. Maybank, and D.~Tao, ``Knowledge distillation: A survey,''
  \emph{International Journal of Computer Vision}, vol. 129, no.~6, pp.
  1789--1819, 2021.

\bibitem{Raffel:19}
C.~Raffel, N.~Shazeer, A.~Roberts, K.~Lee, S.~Narang, M.~Matena, Y.~Zhou,
  W.~Li, and P.~J. Liu, ``Exploring the limits of transfer learning with a
  unified text-to-text transformer,'' 2019.

\bibitem{Milne:06}
D.~Milne, O.~Medelyan, and I.~Witten, ``Mining domain-specific thesauri from
  wikipedia: {A} case study,'' in \emph{Proceedings of the 5th IEEE/WIC/ACM
  International Conference on Web Intelligence (WI 2006 Main Conference
  Proceedings)(WI'06)}, 2006.

\bibitem{Cui:08}
G.~Cui, Q.~Lu, W.~Li, and Y.~Chen, ``Corpus exploitation from {W}ikipedia for
  ontology construction,'' in \emph{Proceedings of the Sixth International
  Conference on Language Resources and Evaluation ({LREC}'08)}.\hskip 1em plus
  0.5em minus 0.4em\relax Marrakech, Morocco: European Language Resources
  Association (ELRA), May 2008.

\bibitem{ferrari:17}
A.~Ferrari, G.~O. Spagnolo, and S.~Gnesi, ``Pure: A dataset of public
  requirements documents,'' in \emph{2017 IEEE 25th International Requirements
  Engineering Conference}, 2017.

\bibitem{Saxena:21}
K.~Saxena, T.~Singh, A.~Patil, S.~Sunkle, and V.~Kulkarni, ``Leveraging
  {W}ikipedia navigational templates for curating domain-specific fuzzy
  conceptual bases,'' in \emph{Proceedings of the Second Workshop on Data
  Science with Human in the Loop: Language Advances}.\hskip 1em plus 0.5em
  minus 0.4em\relax Online: Association for Computational Linguistics, Jun.
  2021, pp. 1--7.

\bibitem{Kluyver:16}
T.~Kluyver, B.~Ragan-Kelley, F.~P{\'e}rez, B.~Granger, M.~Bussonnier,
  J.~Frederic, K.~Kelley, J.~Hamrick, J.~Grout, S.~Corlay, P.~Ivanov, D.~Avila,
  S.~Abdalla, and C.~Willing, ``Jupyter notebooks -- a publishing format for
  reproducible computational workflows,'' in \emph{Positioning and Power in
  Academic Publishing: Players, Agents and Agendas}, 2016.

\bibitem{transformers}
T.~Wolf, L.~Debut, V.~Sanh, J.~Chaumond, C.~Delangue, A.~Moi, P.~Cistac,
  T.~Rault, R.~Louf, M.~Funtowicz, J.~Davison, S.~Shleifer, P.~von Platen,
  C.~Ma, Y.~Jernite, J.~Plu, C.~Xu, T.~L. Scao, S.~Gugger, M.~Drame, Q.~Lhoest,
  and A.~M. Rush, ``Transformers: State-of-the-art natural language
  processing,'' in \emph{Proceedings of the 2020 Conference on Empirical
  Methods in Natural Language Processing: System Demonstrations}.\hskip 1em
  plus 0.5em minus 0.4em\relax Association for Computational Linguistics, 2020.

\bibitem{scikit-learn}
F.~Pedregosa, G.~Varoquaux, A.~Gramfort, V.~Michel, B.~Thirion, O.~Grisel,
  M.~Blondel, P.~Prettenhofer, R.~Weiss, V.~Dubourg \emph{et~al.},
  ``Scikit-learn: Machine learning in {P}ython,'' \emph{Journal of Machine
  Learning Research}, vol.~12, pp. 2825--2830, 2011.

\bibitem{rank-bm25}
\BIBentryALTinterwordspacing
B.~Dorian, J.~Sarthak, N.~Vít, and nlp4whp, ``dorianbrown/rank\_bm25,'' 2022.
  [Online]. Available: \url{https://doi.org/10.5281/zenodo.6106156}
\BIBentrySTDinterwordspacing

\bibitem{beir}
N.~Thakur, N.~Reimers, A.~R{\"u}ckl{\'e}, A.~Srivastava, and I.~Gurevych,
  ``{BEIR}: A heterogeneous benchmark for zero-shot evaluation of information
  retrieval models,'' in \emph{Thirty-fifth Conference on Neural Information
  Processing Systems Datasets and Benchmarks Track (Round 2)}, 2021.

\bibitem{huggingface}
\BIBentryALTinterwordspacing
``Hugging face,'' 2022. [Online]. Available: \url{https://huggingface.co/}
\BIBentrySTDinterwordspacing

\bibitem{wikipy}
\BIBentryALTinterwordspacing
J.~Goldsmith, ``The wikipedia libray,'' 2022. [Online]. Available:
  \url{https://pypi.org/project/wikipedia/}
\BIBentrySTDinterwordspacing

\bibitem{NLTK}
E.~Loper and S.~Bird, ``Nltk: The natural language toolkit,'' in
  \emph{Proceedings of the ACL-02 Workshop on Effective Tools and Methodologies
  for Teaching Natural Language Processing and Computational Linguistics},
  2002.

\bibitem{Du:17}
X.~Du and C.~Cardie, ``Identifying where to focus in reading comprehension for
  neural question generation,'' in \emph{Proceedings of the 2017 Conference on
  Empirical Methods in Natural Language Processing}.\hskip 1em plus 0.5em minus
  0.4em\relax Copenhagen, Denmark: Association for Computational Linguistics,
  2017, pp. 2067--2073.

\bibitem{Pan:19}
L.~Pan, W.~Lei, T.-S. Chua, and M.-Y. Kan, ``Recent advances in neural question
  generation,'' \emph{arXiv preprint arXiv:1905.08949}, 2019.

\bibitem{Miller:95}
G.~Miller, ``{WordNet:} {A} lexical database for {English},''
  \emph{Communications of the ACM}, vol.~38, no.~11, 1995.

\bibitem{Hanna:21}
M.~Hanna and O.~Bojar, ``A fine-grained analysis of {BERTS}core,'' in
  \emph{Proceedings of the Sixth Conference on Machine Translation}.\hskip 1em
  plus 0.5em minus 0.4em\relax Online: Association for Computational
  Linguistics, 2021, pp. 507--517.

\bibitem{Ramage:09}
D.~Ramage, A.~N. Rafferty, and C.~D. Manning, ``Random walks for text semantic
  similarity,'' in \emph{Proceedings of the 2009 workshop on graph-based
  methods for natural language processing (TextGraphs-4)}, 2009, pp. 23--31.

\bibitem{Cambazoglu:21}
B.~B. Cambazoglu, M.~Sanderson, F.~Scholer, and B.~Croft, ``A review of public
  datasets in question answering research,'' in \emph{ACM SIGIR Forum},
  vol.~54, no.~2.\hskip 1em plus 0.5em minus 0.4em\relax ACM New York, NY, USA,
  2021, pp. 1--23.

\bibitem{Whissell:11}
J.~S. Whissell and C.~L. Clarke, ``Improving document clustering using okapi
  bm25 feature weighting,'' \emph{Information retrieval}, vol.~14, no.~5, pp.
  466--487, 2011.

\bibitem{Risch:21}
J.~Risch, T.~M{\"o}ller, J.~Gutsch, and M.~Pietsch, ``Semantic answer
  similarity for evaluating question answering models,'' \emph{arXiv preprint
  arXiv:2108.06130}, 2021.

\bibitem{Sleimi:19}
A.~Sleimi, M.~Ceci, N.~Sannier, M.~Sabetzadeh, L.~Briand, and J.~Dann, ``A
  query system for extracting requirements-related information from legal
  texts,'' in \emph{27th {IEEE} International Requirements Engineering
  Conference}.\hskip 1em plus 0.5em minus 0.4em\relax {IEEE}, 2019.

\bibitem{Kanchev:17}
G.~M. Kanchev, P.~K. Murukannaiah, A.~K. Chopra, and P.~Sawyer, ``Canary: an
  interactive and query-based approach to extract requirements from online
  forums,'' in \emph{2017 IEEE 25th International Requirements Engineering
  Conference}.\hskip 1em plus 0.5em minus 0.4em\relax IEEE, 2017, pp. 470--471.

\bibitem{Hao:22}
T.~Hao, X.~Li, Y.~He, F.~L. Wang, and Y.~Qu, ``Recent progress in leveraging
  deep learning methods for question answering,'' \emph{Neural Computing and
  Applications}, pp. 1--19, 2022.

\bibitem{Yusuf:22}
A.~A. Yusuf, F.~Chong, and M.~Xianling, ``An analysis of graph convolutional
  networks and recent datasets for visual question answering,''
  \emph{Artificial Intelligence Review}, pp. 1--24, 2022.

\bibitem{Jin:19}
H.~Jin, Y.~Luo, C.~Gao, X.~Tang, and P.~Yuan, ``Comqa: Question answering over
  knowledge base via semantic matching,'' \emph{IEEE Access}, vol.~7, pp.
  75\,235--75\,246, 2019.

\bibitem{Diefenbach:20}
D.~Diefenbach, A.~Both, K.~Singh, and P.~Maret, ``Towards a question answering
  system over the semantic web,'' \emph{Semantic Web}, vol.~11, no.~3, pp.
  421--439, 2020.

\bibitem{Ojokoh:18}
B.~Ojokoh and E.~Adebisi, ``A review of question answering systems,''
  \emph{Journal of Web Engineering}, vol.~17, no.~8, pp. 717--758, 2018.

\bibitem{Jing:19}
L.~Jing, C.~Gulcehre, J.~Peurifoy, Y.~Shen, M.~Tegmark, M.~Soljacic, and
  Y.~Bengio, ``Gated orthogonal recurrent units: On learning to forget,''
  \emph{Neural computation}, vol.~31, no.~4, pp. 765--783, 2019.

\bibitem{Wulamu:19}
A.~Wulamu, Z.~Sun, Y.~Xie, C.~Xu, and A.~Yang, ``An improved end-to-end memory
  network for qa tasks,'' \emph{CMC-COMPUTERS MATERIALS \& CONTINUA}, vol.~60,
  no.~3, pp. 1283--1295, 2019.

\bibitem{Ren:20}
Q.~Ren, X.~Cheng, and S.~Su, ``Multi-task learning with generative adversarial
  training for multi-passage machine reading comprehension,'' in
  \emph{Proceedings of the AAAI Conference on Artificial Intelligence},
  vol.~34, no.~05, 2020, pp. 8705--8712.

\bibitem{Parshakova:19}
T.~Parshakova, F.~Rameau, A.~Serdega, I.~S. Kweon, and D.-S. Kim, ``Latent
  question interpretation through variational adaptation,'' \emph{IEEE/ACM
  Transactions on Audio, Speech, and Language Processing}, vol.~27, no.~11, pp.
  1713--1724, 2019.

\bibitem{Wang:18}
A.~Wang, A.~Singh, J.~Michael, F.~Hill, O.~Levy, and S.~R. Bowman, ``Glue: A
  multi-task benchmark and analysis platform for natural language
  understanding,'' \emph{arXiv preprint arXiv:1804.07461}, 2018.

\bibitem{Kumar:19}
V.~Kumar, Y.~Hua, G.~Ramakrishnan, G.~Qi, L.~Gao, and Y.-F. Li,
  ``Difficulty-controllable multi-hop question generation from knowledge
  graphs,'' in \emph{International Semantic Web Conference}.\hskip 1em plus
  0.5em minus 0.4em\relax Springer, 2019, pp. 382--398.

\bibitem{Liu:21}
N.~F. Liu, T.~Lee, R.~Jia, and P.~Liang, ``Can small and synthetic benchmarks
  drive modeling innovation? a retrospective study of question answering
  modeling approaches,'' \emph{arXiv preprint arXiv:2102.01065}, 2021.

\bibitem{Bartolo:21}
M.~Bartolo, T.~Thrush, R.~Jia, S.~Riedel, P.~Stenetorp, and D.~Kiela,
  ``Improving question answering model robustness with synthetic adversarial
  data generation,'' \emph{arXiv preprint arXiv:2104.08678}, 2021.

\bibitem{Lelkes:21}
A.~D. Lelkes, V.~Q. Tran, and C.~Yu, ``Quiz-style question generation for news
  stories,'' in \emph{Proceedings of the Web Conference 2021}, 2021, pp.
  2501--2511.

\bibitem{Gupta:22}
S.~Gupta, A.~Agarwal, M.~Gaur, K.~Roy, V.~Narayanan, P.~Kumaraguru, and
  A.~Sheth, ``Learning to automate follow-up question generation using process
  knowledge for depression triage on reddit posts,'' \emph{arXiv preprint
  arXiv:2205.13884}, 2022.

\end{thebibliography}
